\documentclass[10pt,journal,compsoc]{IEEEtran}

\ifCLASSOPTIONcompsoc
  \usepackage[nocompress]{cite}
\else
  \usepackage{cite}
\fi

\usepackage{times}
\usepackage{epsfig}
\usepackage{graphicx}
\usepackage{amsmath}
\usepackage{amssymb}
\usepackage{sidecap}
\usepackage{wrapfig}
\usepackage{tabu}

\usepackage{array}
\usepackage{epsf}
\usepackage{bm}
\usepackage{latexsym}
\usepackage{float}
\usepackage{subcaption}
\usepackage{color}
\usepackage{algpseudocode}
\usepackage{algorithm}

\graphicspath{{images/}}

\newcommand{\RR}{\mathbb{R}}
\newcommand{\MM}{\cal{M}}

\RequirePackage{amsopn}
\RequirePackage{amsfonts}
\RequirePackage{amsthm}

\begin{document}

\title{ Efficient Inter-Geodesic Distance Computation \\
 		and Fast Classical Scaling}

\author{Gil~Shamai,
        Michael~Zibulevsky,
        and~Ron~Kimmel
\IEEEcompsocitemizethanks{\IEEEcompsocthanksitem Computer Science Department, 
Technion---Israel Institute of Technology, Haifa 3200003.
}}

\IEEEtitleabstractindextext{%
\begin{abstract}
Multidimensional scaling (MDS) is a dimensionality reduction tool used for information analysis, data visualization and manifold learning.
Most MDS procedures embed data points in low-dimensional Euclidean (flat) domains,  such that distances between the points are as close as possible to given inter-point  dissimilarities.
We present an efficient solver for classical scaling, a specific MDS model, by extrapolating the information provided by distances measured from a subset of the points to the remainder.
The computational and space complexities of the new MDS methods are thereby reduced from quadratic to quasi-linear in the number of data points.
Incorporating both local and global information about the data allows us to construct a low-rank approximation of the inter-geodesic distances between the data points.
As a by-product, the proposed method allows for efficient computation of geodesic distances.
\end{abstract}
}

\maketitle

\IEEEdisplaynontitleabstractindextext

%
\IEEEpeerreviewmaketitle

\IEEEraisesectionheading{\section{Introduction}\label{sec:introduction}}

\IEEEPARstart{W}{ith}
the increasing availability of digital information, the need for data simplification and dimensionality reduction tools constantly grows.
\textit{Self-Organizing Map} (SOM) \cite{kohonen1998self}, \textit{Local Coordinate Coding} \cite{yu2009nonlinear,zhou2013locality}, \textit{Multidimensional Scaling} (MDS), and \textit{Isomap} \cite{borg2005modern, tenenbaum2000global} are examples of data reduction techniques that are used to simplify data and hence reduce computational and space complexities of related procedures. 

MDS embeds the data into a low-dimensional Euclidean (flat embedding)
 space, while attempting to preserve the distance between each pair 
 of points. 
Flat embedding is a fundamental step in various applications
 in the fields of data mining, statistics, manifold learning, non-rigid
 shape analysis, and more.
For example, in \cite{Panozzo:2013aa}, Panozzo et al. demonstrated, using MDS, how to generalize Euclidean weighted averages to weighted averages on surfaces. 
Weighted averages on surfaces could then be used to efficiently compute splines on surfaces, solve re-meshing problems, and find shape correspondences.
Zigelman et al. \cite{zigelman2002texture} used MDS for texture mapping, 
 while Schwartz et al.  \cite{schwartz1989numerical,drury1996computerized}
 utilized it to flatten models of monkeys' cortical surfaces for further analysis.
In \cite{schweitzer2001template,rubner2000perceptual,pless2003image}, 
 MDS was applied to image and video analysis. 
In \cite{elad2003bending} and \cite{shamai2017geodesic}, MDS was used to construct a bending invariant representation of surfaces, referred to as a {\it canonical form}. 
Such canonical forms reveal the intrinsic structure of the surface, and were used to simplify tasks such as non-rigid object matching and classification.

The bottleneck step in MDS is computing and storing all pairwise distances.
For example, when dealing with geodesic distances computed on surfaces, 
 this step is time-consuming and, in some cases, impractical. 
The \textit{fast marching} method \cite{kimmel1998computing} is an example of one efficient procedure for approximating the geodesic distance map between all pairs of $p$ points on a surface.
Its complexity is $O(p^2 \log p)$ when used in a straightforward manner. 
Nevertheless, even when using such an efficient procedure, the complexities involved in computing and storing these distances are at least quadratic in the number of points.

One way to reduce these complexities is by considering only local
 distances between nearby data points. 
Attempts to do so were made, for example, in \textit{Locally Linear Embedding} (LLE,  \cite{roweis2000nonlinear}) \textit{Hessian Locally Linear Embedding} (HLLE, \cite{donoho2003hessian}), and Laplacian eigenmaps \cite{belkin2001laplacian}. 
For these methods, the effective time and space complexities are linear in the number of points. 
These savings, however, come at the expense of only capturing the local structure while the global geometry is ignored.
In \cite{silva2002global}, De Silva and Tenenbaum suggested computing only distances between a small subset of landmarks.
The landmarks were then embedded, ignoring the rest of the distances, and then the rest of the points were interpolated into the target flat space. 
The final embedding, however, was restricted to the initial one at the landmark points.
 
The Nystr{\"o}m method \cite{arcolano2010Nystrom} is an efficient technique
 to reconstruct a positive semidefinite matrix using only a small subset 
 of its columns and rows. 
Various recent techniques for approximating and reducing the complexities of MDS bear some resemblance to the Nystr{\"o}m method \cite{platt2005fastmap}.
In \cite{yu2012isomap} and \cite{civril2007ssde}, only a few columns of the pairwise distance matrix were computed, after which the Nystr{\"o}m method was used to interpolate the rest of the matrix. 
The main differences between these two methods are the column sampling 
 schemes and the way to compute the initial geodesic distances. 
In the latter method, a regularization term was used for the pseudo-inverse in the Nystr{\"o}m method, which further improved the distance matrix reconstruction.
\cite{liu2006sub} combined the Nystr{\"o}m method with Kernel-PCA
 \cite{scholkopf1998nonlinear} and demonstrated an efficient application 
 of mesh segmentation.
Recently, Spectral-MDS (SMDS) \cite{aflalo2013spectral} obtained 
 state-of-the-art results for efficiently approximating the embedding 
 of MDS. 
There, geodesic distances are interpolated by exploiting a smoothness assumption through the Dirichlet energy. 
Complexities are reduced by translating the problem into the spectral domain and representing all inter-geodesic distances, considering only the first eigenvectors of the Laplace-Beltrami operator (LBO).

Here, we explore two methods, the Fast-MDS (FMDS) \cite{shamai2015classical} and Nystr{\"o}m-MDS (NMDS) \cite{shamai2015accelerating}, for efficiently approximating inter-geodesic distances and, thereby, reduce the complexities of MDS.
FMDS interpolates the distance map from a small subset of landmarks using 
 a smoothness assumption as a prior, formulated through the bi-Laplacian operator \cite{lipman2010biharmonic, botsch2008linear, stein2018natural}.
As opposed to Spectral-MDS, the problem is solved in the spatial domain and no eigenvectors are omitted, so that accuracy is improved. Nevertheless, the time complexities remain the same.
NMDS learns the distance interpolation coefficients from an initial set of computed distances.
Both methods reconstruct the pairwise geodesic distance matrix as a low-rank product of small matrices and obtain a closed form approximation of \textit{classical scaling} through small matrices.

Our numerical experiments compare FMDS and NMDS to all relevant methods mentioned above  and demonstrate high efficiency and accuracy in approximating the embedding of MDS and the inter-geodesic distances.
In practice, our methods embed $10K$-vertex shapes within less than a second, including all initializations, with an approximation error of $0.007\%$ from the same embedding obtained by MDS.
When compared to \textit{exact geodesics} \cite{surazhsky2005fast} on surfaces, our methods approximate geodesic distances with a better average accuracy than fast marching, while computing $500$M geodesic distances per second.

The paper is organized as follows.
In Section \ref{sec:review_MDS} we briefly review the classical scaling
 method.
In Section \ref{sec:Distances_Interpolation} we develop two methods 
 for geodesic distance interpolation from a few samples, formulated 
 as a product of small-sized matrices.
Section \ref{sec:rows_selection} discusses how to select the samples.
In Section \ref{sec:acceleration} we reformulate classical 
 scaling through the small-sized matrices obtained by the interpolation.
Section \ref{sec:Experimental_Results} provides support for the proposed
 methods by presenting experimental results and comparing to other methods.
Finally, Section \ref{sec:applications} discusses 
 an extension of embedding on a sphere.

\section{Classical Scaling}
\label{sec:review_MDS}
Given $p$ points $\{y_i\}_{i=1}^p$ equipped with some similarity measures
 $D_{ij}$ 
 between them, multidimensional scaling (MDS) methods aim at finding an
 embedding $\{z_i\}_{i=1}^p$ in a low-dimensional Euclidean space $\RR^m$  
 such that the  Euclidean distances $\left\| {{z_i} - {z_j}} \right\|_{\RR^m}$
 are as close as possible to $D_{ij}$.
When the points $\{y_i\}_{i=1}^p$ lie on a manifold, the affinities $D_{ij}$ can 
 be defined as the geodesic distances between them.
In this case, on which we focus in this paper, $D_{ij}$ are invariant to isometric
 deformations of the surface, thus revealing its intrinsic geometry.
One way to define this problem, termed classical scaling, is through the minimization
  \begin{equation}
  \arg_Z \min {\left\| {Z{Z^T} + \frac{1}{2}JEJ} \right\|_F},
  \label{eq:MDS}
  \end{equation}
   where $Z_{ij}=z_i^j$,  ${E_{ij}} = D_{ij}^2$ and ${J_{ij}} = {\delta _{ij}} - \frac{1}{p}$, and
   as usual, $\delta _{ij} = 0$ for $i \neq j$ and $\delta _{ii} = 1$ for all $i$.
Denoting by $\tilde {V}\tilde {\Lambda} \tilde {V}^T$ the thin eigenvalue
   decomposition of the symmetric matrix $-\frac{1}{2}JEJ$, with only the $m$ 
   largest eigenvalues and corresponding eigenvectors, the solution for this 
   problem is given by $Z = \tilde {V}\tilde{\Lambda}^{\frac{1}{2}}$.  
This requires the computation of all pairwise distances in the $p\times p$ 
 matrix $E$, which is not a practical solution method when dealing with more than 
 a few thousand points.
In this paper, we reconstruct $E$ from a small set of its columns. 
We formulate the reconstruction as a product of small-sized matrices, 
 and show how to solve the above minimization problem without explicitly computing 
 the reconstruction. 
Thus, we circumvent the bottleneck of classical scaling and  significantly reduce both space and time complexities from quadratic to quasi-linear.

\section{Distance Interpolation}
\label{sec:Distances_Interpolation}

Let $\MM$ be a manifold embedded in some $\RR^k$ space.
Denote by $E(x,y)$ the squared geodesic distance between $x,y\in \MM$.
In the discrete domain $\MM$ is represented by $p$ points, and $E$ is represented by a $p \times p$ matrix.
In Sections \ref{sec:FMDS_decomposition} and \ref{sec:NMDS_decomposition} we develop two methods for reconstructing $E$ by decomposing it 
 into smaller-sized matrices $E\approx STS^T$.

\subsection{A smoothness-based reconstruction}
\label{sec:FMDS_decomposition}
One cornerstone of numerical algebra is representing a large matrix as a
 product of small-sized matrices
  \cite{drineas2006fast,mahoney2009cur}.
Compared to such classical decomposition, better approximations can 
 be obtained  when considering a specific data model.
Here, we exploit the fact that the elements of the matrix to be reconstructed 
 are squared geodesic distances, derived from some smooth manifold.

\begin{figure}[htbp]
\begin{center}
\includegraphics[width=1\columnwidth]{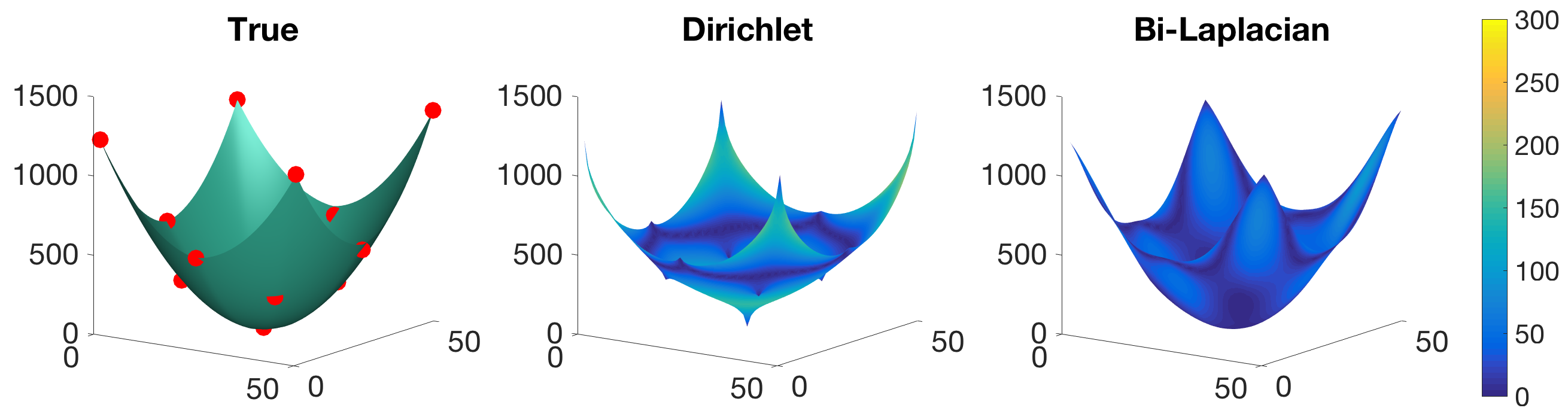}
\end{center}
\caption{Reconstruction of $e$ (a column of $E$) on a flat surface.
Left: The true values of $e$. The chosen $n=13$ samples
are marked by red points. 
Middle and right:
The reconstructed function $e^*=Mr$ using the Dirichlet and Laplacian energies. For comparison,
we colored the function according to the absolute error $|e^*-e|$.}
\label{fig:flatD}
\end{figure}

Starting from the continuous case, 
let $x_0 \in \MM$ be an arbitrary point on the manifold, and denote by 
 $\bar e  \colon x \in \MM \to \RR$ the squared geodesic distance from $x_0$ 
 to all other points $x \in \MM$. 
Assume the values of $\bar e(x)$ are known in a set of $n$ samples 
 $\{x_i\}_{i=1}^n \in \MM$, and denote these values by $\{r_i\}_{i=1}^n$.
We expect the distance function $\bar e(x)$ to be smooth on the manifold, 
 as nearby points should have similar $\bar e(x)$ values.
Therefore, we aim to find a function $\bar e(x)$ that both is smooth 
 and satisfies the constraints $\bar e(x_i)=r_i$.
This can be formulated through the minimization
 \begin{equation}
 \arg_{\bar e} \min{\mathcal{E}(\bar e)} \hspace{5 mm} \mbox{s.t.} \hspace{5 mm}
 \bar e(x_i)=r_i,
 \label{eq:energy_minimization}
  \end{equation}
 where the energy $\mathcal{E}(\bar e)$ is some smoothness measure of 
 $\bar e(x)$ on $\MM$.
One possible, somewhat sensitive, choice of $\mathcal{E}(\bar e)$ is 
 the Dirichlet energy
\begin{equation}
\mathcal{E}(\bar e) = \int\limits_{x \in \MM} {\| \nabla \bar e(x) \|^2_2da(x)},
\label{eq:energy_dirichlet}
\end{equation}
 used in \cite{aflalo2013spectral}, where $da(x)$ is the infinitesimal volume 
  element and the gradient is computed with respect to the manifold.
  The Dirichlet energy with point constraints could lead to removable discontinuities;
  see Figure \ref{fig:flatD}.
Therefore, instead, we use a different popular measure 
 that yields a decomposition that is both simple and less sensitive.
Denoting by $\bar l(x) = \Delta \bar e(x)$ the result of the LBO on $\MM$ applied to $\bar e$, this energy is defined as 
\begin{equation}
 \mathcal{E}(\bar e) = \int\limits_{x \in \MM} {{{\left( \Delta \bar e(x) \right)}^2}da(x)}
      = \int\limits_{x \in \MM} {{{\left( {\bar l(x)} \right)}^2}da(x)}.
       \label{eq:energy_laplacian}
\end{equation}
It yields the bi-Laplacian as the main operator in the resulting Euler-Largange equation \cite{lipman2010biharmonic, botsch2008linear, stein2018natural}.

\begin{figure}[htbp]
\begin{center}
\includegraphics[width=1\columnwidth]{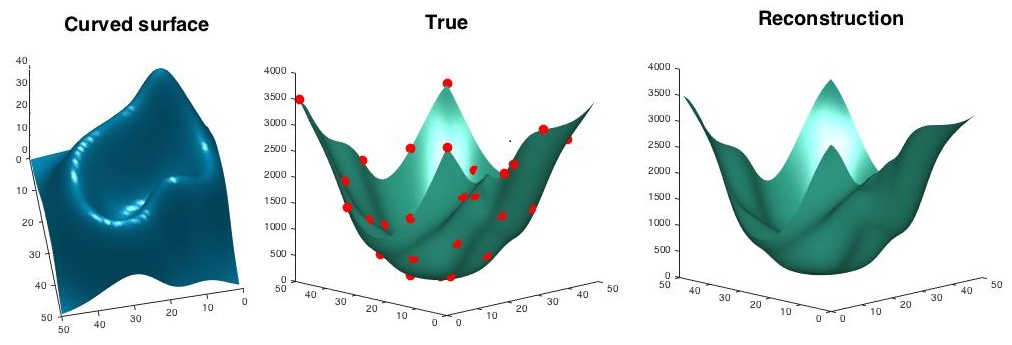}
\end{center}
\caption{Reconstruction of $e$ (a column of $E$) on a curved surface. Left: The curved surface.
Middle: The true values of the distance function $e$, measured from the middle point of the surface, and sampled at $n=30$ red points. 
Right: The reconstructed function $e^*=Mr$ using the suggested energy.}
\label{fig:curveD}
\end{figure}

In the discrete case, 
 we define the diagonal matrix $A$ such that its diagonal 
 is a discretization of $da(x)$ about each corresponding vertex,
 and denote by the vectors $l$ and $e$ the discretization of $\bar l$ and 
 $\bar e$, respectively.
Note that $e$ is a column in the matrix $E$, 
 which was defined earlier as the pairwise squared geodesics.
Denote by $L$ the discretization of the LBO, such that $l=Le$.
We use the cotangent-weight Laplacian with Dirichlet boundary conditions in our experiments \cite{pinkall1993computing}, \cite{belkin2001laplacian}. 
Any discretization matrix of the LBO, however, can be used. 
Following these notations, the energy now reads
\begin{equation}
 \mathcal{E}(e) = \sum\limits_{i = 1}^p {{l_i}^2{A_{ii}}}  
                = {l^T}Al = {e^T}{L^T}ALe,
\end{equation}
and the interpolation becomes
\begin{equation}
 e^* = \arg_e \min {e^T}{L^T}ALe \hspace{5 mm} 
  \mbox{s.t.} \hspace{5 mm} Be=r,
\end{equation}
 where $r$ is an $n \times 1$ vector holding the values $\{r_i\}_{i=1}^n$. 
Recall that $r$ is a subset of $e$. 
Hence, the $n\times p$ matrix $B$ can easily be defined such that $Be=r$.
Since the constraints may be contaminated with noise, 
 a relaxed form of the problem using a penalty function is used instead. 
This can be formulated as
\begin{equation}
e^* = \mathop {\arg_e \min } ({e^T}{L^T}ALe + \mu {\left\| {Be - r} \right\|^2}),
\label{eq:pen}
\end{equation}
where $\mu$ is a sufficiently large scalar.
 This is a quadratic equation and its solution is given by
\begin{eqnarray}
\label{eq:M}
M &=& {({L^T}AL + \mu {B^T}B)^{ - 1}}\mu {B^T}\cr
e^*  &=& Mr.
\end{eqnarray}
To sum up, given a set of samples $r$ of the distance function $e$, 
 $Mr$ is a reconstruction of $e$ in the sense of being a smooth interpolation 
 from its samples.
Figures \ref{fig:flatD} and \ref{fig:curveD} demonstrate the reconstruction 
 of $e$ from its samples for flat and curved manifolds. 
Recall that $e$ is the squared geodesic distance from a point $x_0$ to the rest 
 of the points on the manifold. 
In this example, we chose $x_0$ at the center of the manifold $\MM$.
In Figure \ref{fig:flatD}, the surface is flat and hence the function $e$ is simply the squared Euclidean distance from a point in $\RR^2$, $z=x^2+y^2$,
where $x$ and $y$ are the Euclidean coordinates.
We compare the suggested energy to the Dirichlet energy. 
As can be seen, when using the Dirichlet energy, the function includes sharp discontinuities at the constraint points.

In Equation (\ref{eq:pen}), a larger value of $\mu$ corresponds to stronger constraints, in the sense that the surface is restricted to pass closer to the constraints. 
In the next experiment, we measure the average reconstruction error 
$\frac{\|e^*-e\|_2}{\|e\|_2}$,
with respect to the parameter $\mu$, for a random set of geodesic paths $\{e\}$ on the hand and giraffe shapes from the SHREC dataset. 
It can be seen that larger values of $\mu$ lead to better approximations up to a limit that depends on the number of samples. 
This saturation occurs at a point where the surface approaches the constraints up to a negligible distance with respect to its reconstruction error at the remaining points. 
Very large values of $\mu$ might lead to numerical errors in the inversion step of Equation (\ref{eq:M}). 
In practice, as a typical number of samples in our method is $20$--$100$, any value between $10^4$ and $10^{10}$ would be a good choice of $\mu$.
\begin{figure}[htbp]
\centering
\includegraphics[width=0.49\columnwidth]{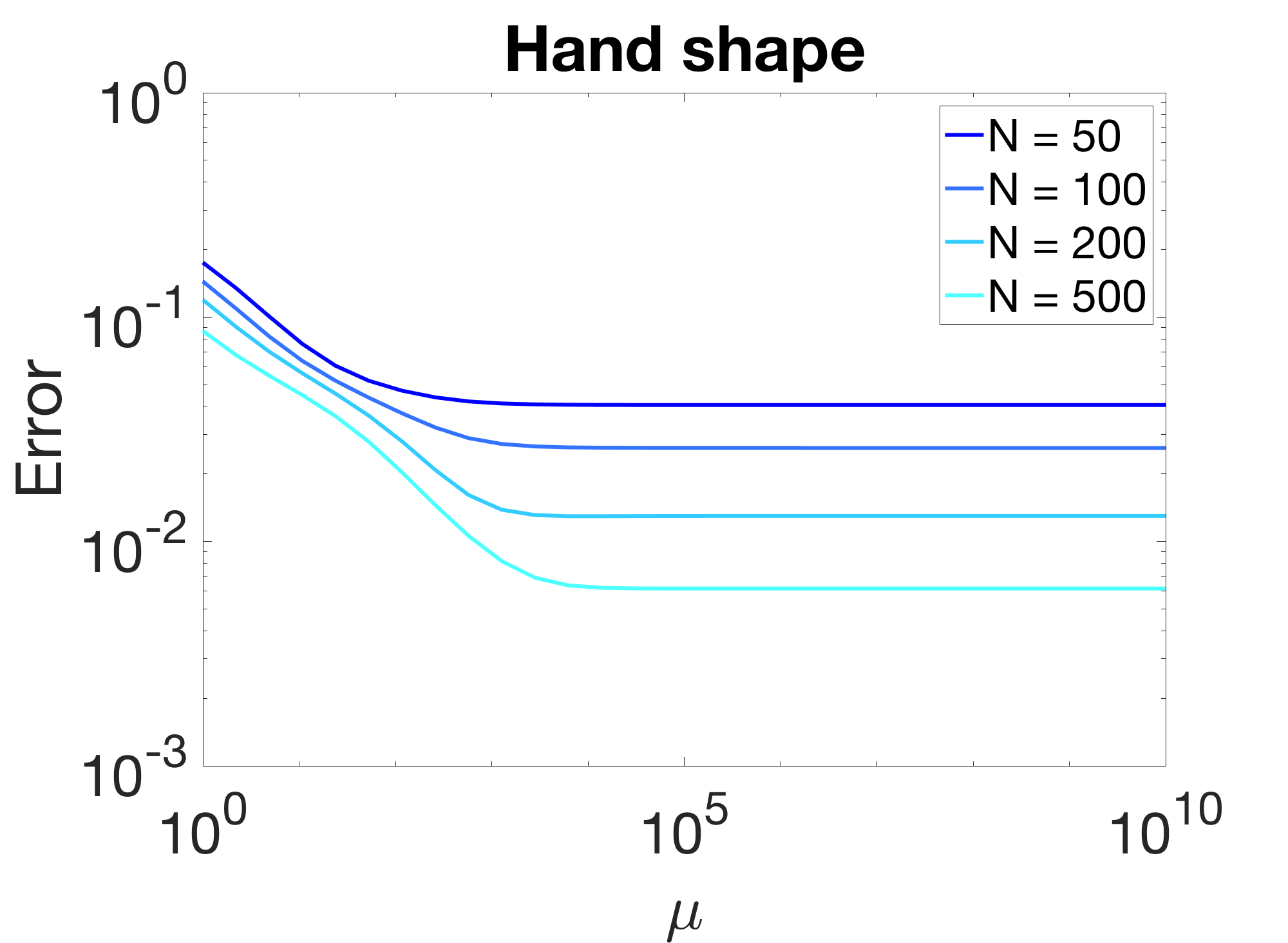} 
\includegraphics[width=0.49\columnwidth]{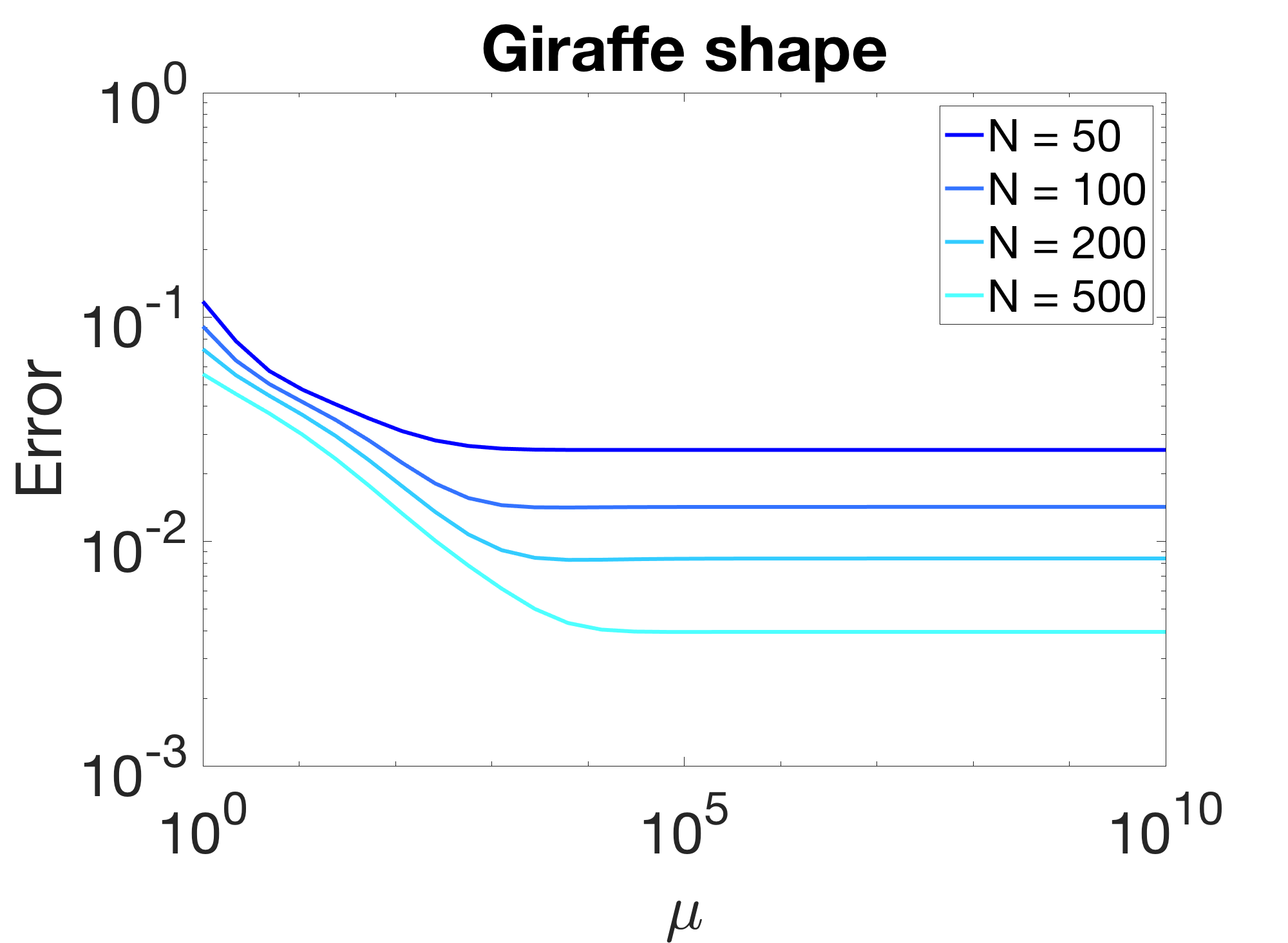} 
\caption{Average reconstruction error w.r.t. the penalty parameter $\mu$.}
\label{fig:results_mu}
\end{figure}

Next, since $e$ is a column in the matrix $E$, we reconstruct all the other columns in the same manner. 
Notice that the matrix $M$ does not depend on the values of $e$, but only
on the samples' locations. 
Hence, it can be computed once and used to reconstruct all columns. 
Moreover, it is possible to formulate the reconstruction of all the columns of $E$ simultaneously using a simple matrix product.
Let us choose $n \ll p$ columns of $E$ and let the $p \times n$ matrix $R$ 
 hold these columns. 
The locations of the columns in $E$ correspond to the locations of 
 the samples in the manifold. 
Each column of $R^T$ corresponds to a vector $r$, which contains the distances
 from the $n$ samples to a point $x_0$, which corresponds to the column.
The column selection and the computation of $R$ are discussed in Section \ref{sec:rows_selection}.
Now, all columns of $E$ can be interpolated simultaneously by the product
\begin{equation}
\hat E = MR^T.
\end{equation}
Since $E$ is symmetric by definition, we symmetrize its
  reconstruction by
\begin{equation}
\hat E = \frac{1}{2}(MR^T + RM^T).
\label{eq:rec}
\end{equation}
Notice that $\hat E$ is a low-rank reconstruction of $E$ with a maximal 
 rank of $2n$, because $R$ is of size $p \times n$, which means the rank of $MR^T$ is at most $n$. Therefore,
the rank of $\hat E = \frac{1}{2}(MR^T + RM^T)$ is at most $2n$. 
Consequently, $\hat E$ can be written as a product of matrices that are no larger than $2n \times p$.
These matrices can be obtained as follows.
Denote by ${S = (M|R)}$ the horizontal concatenation of $M$ and $R$, and define the block matrix $T = \frac{1}{2}\left( {\begin{array}{*{20}{c}} 
   {{0_{n\times n}}}&{{I_{n\times n}}}\\
   {{I_{n\times n}}}&{{0_{n\times n}}}
   \end{array}} \right)$,
 where $I$ is the identity matrix.
It can be verified that
\begin{equation}
 \hat E = \frac{1}{2}(MR^T + RM^T) =ST{S^T}.
\label{eq:equation3}
\end{equation}
We actually only need to keep the matrices $S$ and $T$ instead of the whole matrix $\hat E$, which reduces the space complexity from $O(p^2)$ to $O(np)$.
In our experiments, between $20$ and $100$ samples were enough for accurate reconstruction of $E$, where the number of vertices $p$ was in the range of $10^3$ to $10^6$.

Above we derived a low-rank approximation of the matrix $E$ from 
   a set of known columns. 
The same interpolation ideas can be used for other tasks such as 
 geodesics extrapolation \cite{campen2011walking} and matrix completion \cite{fang2012euclidean}, in which a matrix is reconstructed from a set of known 
 entries and the columns may vary in their structure.

\subsection{A learning-based reconstruction}
\label{sec:NMDS_decomposition}
In Section \ref{sec:FMDS_decomposition}, we derived a reconstruction $\hat E = STS^T$ by
exploiting the smoothness assumption of the distances on the manifold. 
Although the suggested energy minimization results in a simple low-rank reconstruction of $E$, this might not be the best way to interpolate the distances.
We now formulate a different decomposition based on a learning approach.

\begin{figure}[htbp]
\begin{subfigure}[t]{0.2\textwidth}
\includegraphics[width=\linewidth]{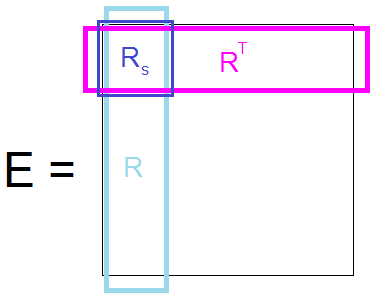}
\caption{}
\label{fig:matrixPartition1}
\end{subfigure}\hfill
\begin{subfigure}[t]{0.2\textwidth}
\includegraphics[width=\linewidth]{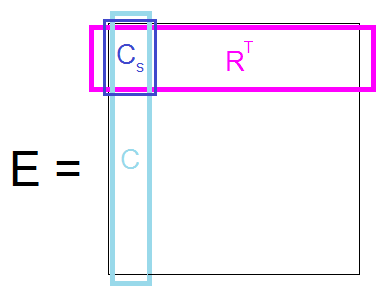}
\caption{}
\label{fig:matrixPartition2}
\end{subfigure}
\caption{Demonstration of the partition of $E$, assuming the columns 
  were chosen as the first ones. 
$R_s$ and $C_s$ are a part of the matrices $R$ and $C$. Note that $C$ and $C_s$ are subsets of $R$ and $R_s$.}
\end{figure}

As before, we choose $n$ arbitrary columns of $E$ and let $R$ hold
these columns. 
In Section \ref{sec:FMDS_decomposition}, we constructed an interpolation matrix $M$ using the LBO as a smoothness measure such that 
$E \approx MR^T$ or similarly $E \approx RM^T$. 
Here we attempt to answer the following question:
Is it possible to construct a matrix $M$ that would yield a better
 reconstruction?
The matrix $M$ for the best reconstruction of $E$ in terms of the Frobenius 
 norm is obtained by 
\begin{equation}
\label{eq:best_reconstruction}
 \arg_M \min \|RM^T - E\|_F^2,
\end{equation}
and the solution is given by $M^T=R^+E$. 
In this case, the reconstruction is nothing but a projection of $E$ 
 onto the subspace spanned by the columns of $R$, 
\begin{equation}
  \hat E = RR^+E.
\label{eq:projection}
\end{equation}
We cannot, however, compute the best $M$ since we do not know the entire matrix $E$.
Hence, we suggest learning the coefficients of $M$ from the part of $E$ that we do know.
This can be formulated using a Hadamard product as
\begin{equation}
 \arg_M \min \|H\circ (RM^T - E)\|_F^2, 
\end{equation}
 where $H$ is a mask of the location of $R^T$ in $E$ 
 (the known rows of $E$). 
 The known rows of $E$ can be thought of as our training data for 
 learning $M$.
 This has some resemblance to various matrix completion formulations, such as the Euclidean Distance Matrix (EDM) completion in \cite{fang2012euclidean}. 
 Nevertheless, unlike previous methods, our primary goal here is to derive a formulation that can be expressed as a low-rank matrix product, preferably in a closed form, which is also efficient to compute.
Next, define $R_s$ as the intersection of $R$ and $R^T$---that is,
 the elements that belong both to $R$ and $R^T$; see Figure \ref{fig:matrixPartition1}. 

The above equation is equivalent to  
\begin{equation}
 \arg_M \min \|R_sM^T - R^T\|_F^2.
\end{equation}
In this form, the number of coefficients in $M$ is large with respect to our training data, which will result in overfitting. For instance, for a typical $R_s$ with independent columns, the reconstruction of the training data $R^T$ is perfect. Therefore, to avoid overfitting to the training data, we regularize this equation by reducing the number of learned coefficients.
Let $C$ hold a subset of $n_1 \leq n$ columns of $R$, and define $C_s$ as the intersection of $C$ and $R^T$; see Figure \ref{fig:matrixPartition2}.
The regularized formulation is 
\begin{equation}
  \arg_M \min \|C_sM^T - R^T\|_F^2,
\end{equation}
where $M$ is now a smaller matrix of size $p \times n_1$, and the solution is given by
\begin{equation}
M = (C_s^TC_s)^{-1}C_s^TR^T = UR^T,
\end{equation}
yielding what is known as a CUR decomposition of $E$, 
\begin{equation}
  \hat E = CUR^T.
  \label{eq:CUR}
\end{equation}
CUR decompositions are widely used and studied; see, 
 for example, \cite{drineas2006fast,mahoney2009cur}.
Notice that when $n_1 = n$, we obtain
\begin{equation}
  \hat E = RR_s^+R^T,
\end{equation}
which is known as the Nystr{\"o}m decomposition for positive semidefinite matrices \cite{arcolano2010Nystrom}.
Choosing  a large value for $n_1$ would result in overfitting to the training data. Alternatively, a low value would limit the rank of the approximated matrix and hence the ability to capture the data. 

\begin{figure}[htbp]
\begin{center}
\includegraphics[width=0.9\columnwidth]{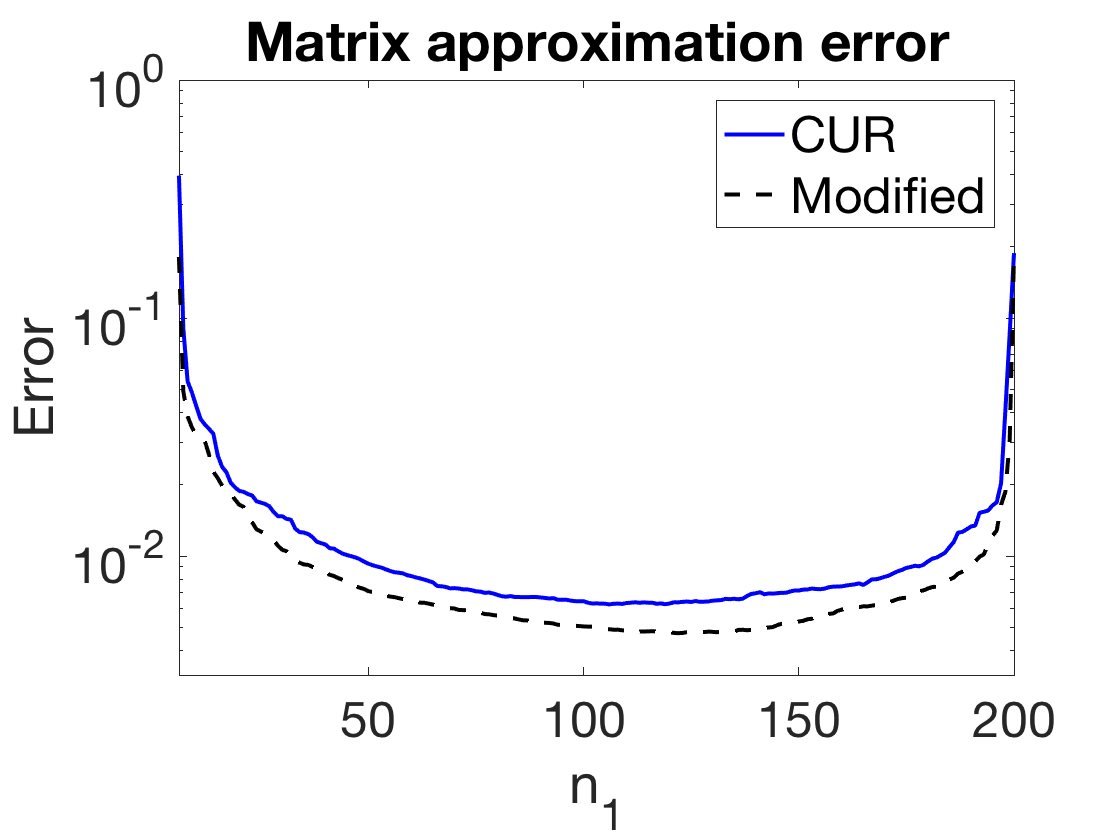}
\end{center}
\caption{The reconstruction error with respect to $n_1$, given by 
$\frac{\|\hat E-E\|_F}{\|E\|_F}$, using the learning
approach (CUR) and its proposed modification. 
 $E$ is computed from the giraffe shape in the SHREC database 
  (see experimental results).}
\label{fig:nystromN1}
\end{figure}

In Figure \ref{fig:nystromN1} (CUR) we demonstrate the reconstruction error of $E$ with respect to $n_1$, where $n = 200$. 
There is a significant potential improvement compared to the previous un-regularized case of $n_1 = n$, which results in the Nystr{\"o}m method. 

In order to further reduce overfitting, we defined the matrix $C$ as a subset 
 of the matrix $R$. 
A different choice of $C$ can lead to a better reconstruction 
 of $E$. 
Denote by $V \Lambda V^T$ the eigenvalue decomposition of $R_s$ and
 denote by $\tilde V \tilde \Lambda \tilde V^T$ the thin eigenvalue  
 decomposition of $R_s$ with only the $n_1$ eigenvalues with the largest
 magnitude and $n_1$ corresponding eigenvectors.
The choice of 
\begin{equation}
C = R\tilde V
\end{equation}
formulates $C$ as a linear combination of the columns of $R$ instead of
 a subset of $R$, thus exploiting better learning of the data. 
Moreover, this will result in a symmetric decomposition of $E$. 
 We have 
 \begin{eqnarray}
C_s &=& R_s \tilde V\cr 
&=& V \Lambda V^T \tilde V \cr
&=& V\Lambda V^TVI_{n\times n_1} \cr
&=& V\Lambda I_{n\times n_1} \cr
&=& \tilde V\tilde \Lambda,
\end{eqnarray}
where $I_{n\times n_1}$ is the rectangular identity matrix (with ones along the main diagonal and zeros elsewhere).
Thus,
\begin{eqnarray}
U &=& (C_s^TC_s)^{-1}C_s^T  \cr
&=& (\tilde \Lambda \tilde V^T \tilde V\tilde \Lambda)^{-1}\tilde \Lambda \tilde V^T \cr &= &
(\tilde \Lambda^2)^{-1}\tilde \Lambda \tilde V^T \cr
&=& \tilde \Lambda^{-1} \tilde V^T.
\end{eqnarray}
Plugging $C = R\tilde V$ and $U = \tilde \Lambda^{-1} \tilde V^T$ 
  into Equation (\ref{eq:CUR}), we obtain
\begin{equation}
  \hat E = R\tilde V \tilde \Lambda^{-1} \tilde V^T R^T.
  \label{eq:NMDS_reg}
\end{equation}
Finally, by defining $S = R$ and $T = \tilde V \tilde \Lambda^{-1} \tilde V^T$,
 we obtain the desired decomposition
\begin{equation}
  \hat E = STS^T.
  \label{eq:sts}
\end{equation}
This results in a symmetric low-rank decomposition of $E$ with rank $n_1$ 
 (compared to the rank $2n$ we achieved in Section \ref{sec:FMDS_decomposition}). 
It can be thought of as a variant of the Nystr{\"o}m method where $T = \tilde V \tilde \Lambda^{-1} \tilde V^T$ is a regularized version of $R_s^+$. 
In Figure \ref{fig:nystromN1} 
 we demonstrate the improvement resulting from this modification.
In all our experiments,  $\left\lceil\frac{1}{2}n\right\rceil$ 
 was a good choice for $n_1$. 

\section{Choosing the set of columns}
\label{sec:rows_selection}
The low-rank approximations of $E$ developed in Section
 \ref{sec:Distances_Interpolation} are constrained to the subspace 
 spanned by chosen columns. 
Thus, a good choice of columns would be one that captures most of the 
 information of $E$. 
The \textit{farthest point sampling strategy} is a 2-optimal
 method for selecting points from a manifold that are far away from each
 other  \cite{hochbaum1985best}. The first point is chosen arbitrarily. Then, recursively, the next point is chosen as the farthest (in a geodesic sense) from the already chosen ones. After each point is selected, the geodesic distance from it to the rest of the points is computed.

 The computation of geodesic distance from one point to the rest of the points can be
performed efficiently using the fast marching method
\cite{kimmel1998computing} for two-dimensional triangulated surfaces,
 and \textit{Dijkstra's shortest path algorithm} \cite{drrksrranote}
 for higher dimensions. 
 For surfaces with $p$ vertices, both methods have complexities 
  of $O(p\log p)$. 
Therefore, the complexity of choosing $n$ points with
 the farthest point sampling strategy is $O(np\log p)$.
 Note that Dijkstra's algorithm might have some limitations as an interpolation mechanism and should be used with care, especially in high-dimensional manifolds \cite{stein2018natural}. 
 For the case of a weighted graph, the authors in \cite{stein2018natural} suggested a way to compute the edge weights for convergence of the graph to the actual distance.
 
Using the farthest point sampling strategy, we obtain
 $n$ samples from the manifold and their distances to the rest
 of the points, which, when squared, correspond to $n$ columns of $E$. 
Since the samples are far from each other, the corresponding columns
 are expected to be far as well (in an $L_2$ sense) and serve as a good
 basis for the image of $E$.
While other column selection methods need to store the whole matrix 
 or at least scan it several times to decide which columns to choose,
 here, by exploiting the fact that columns correspond to geodesic 
 distances on a manifold, we do not need to know
 the entire matrix in advance.
The farthest point sampling method is described in
 Procedure \ref{alg:FPS}.
\begin{algorithm}
\floatname{algorithm}{Procedure}
\caption{Farthest point sampling}
\begin{algorithmic}[1]

\renewcommand{\algorithmicrequire}{\textbf{Input}}
\renewcommand{\algorithmicensure}{\textbf{Output}}
\Require A manifold with a set of $p$ vertices $\mathcal{V} = \{v_1, v_2, ... , v_p\}$ and desired number of chosen vertices $n$.
\Ensure A set of samples $\mathcal{S}=\{r_1, ..., r_n\}$ and their distances to the rest
of the vertices $F_{p\times n}$

\State Choose an initial vertex $r_1$,
$\mathcal{S} \gets \mathcal{S} \cup \{r_1\}$
\State Compute ${F_{(:,1)} \gets \mbox{dist}(v_{r_1})}$ (see comment below).
    \For{$i=2$ to $n$}
        \State Find the farthest vertex from the already chosen ones,
        ${r_i = \arg\max_{1\leq j \leq |V|} \min_{1\leq k < i} F_{jk}}$
        \State Update the set of selected samples, $\mathcal{S} \gets \mathcal{S} \cup \{r_i\}$
        \State Compute ${F_{(:,i)} \gets \mbox{dist}(v_{r_i})}$.
    \EndFor
    \\
\Comment{ $\mbox{dist}(v)$ returns a vector of geodesic distances from 
           vertex $v$ to the rest of the vertices.}
\end{algorithmic}
\label{alg:FPS}
\end{algorithm}

\section{Accelerating Classical Scaling}
\label{sec:acceleration}
In this section, we show how to obtain the solution for 
 classical scaling using the decomposition $\hat E = STS^T$ constructed
 in the previous sections.
A straightforward solution would be to plug $\hat E$ instead of $E$ into Equation \ref{eq:MDS}
 and repeat the steps: 
\begin{enumerate}
    \item Compute the thin eigenvalue decomposition of
 \mbox{$Y = -\frac{1}{2}J\hat E J$} with the $m$ largest eigenvalues and 
 corresponding eigenvectors. 
This can be written as
\begin{equation}
 Y \approx \tilde V_1 \tilde \Lambda_1 {\tilde V_1^T}.
\label{eq:thin_Y}
\end{equation}
    \item Compute the embedding by
\begin{equation}
 Z = \tilde V_1{\tilde \Lambda_1 ^{\frac{1}{2}}}.
\label{eq:equation1}
\end{equation}
\end{enumerate}
This solution, however, requires storing and computing $\hat E$,  resulting in high computational and memory complexities.

To obviate this, we propose the following alternative.
Denote by $Q\mathcal R$ the thin QR factorization of ${JS}$.
Note that the columns of $Q$ are orthonormal and $\mathcal R$ is a small upper triangular matrix.
We get
\begin{equation}
Y = -\frac{1}{2}J\hat EJ = -\frac{1}{2}JST{S^T}J =  -\frac{1}{2}Q\mathcal RT{\mathcal R^T}{Q^T}.
\label{eq:equation4}
\end{equation}
Denote by $\tilde V_2\tilde \Lambda_2 {\tilde V_2^T}$ the thin eigenvalue decomposition of $ -\frac{1}{2}\mathcal RT{\mathcal R^T}$, keeping only the $m$ first eigenvectors and $m$ corresponding largest eigenvalues. 
This can be written as
\begin{equation}
  -\frac{1}{2}\mathcal RT{\mathcal R^T} \approx \tilde V_2\tilde \Lambda_2 {\tilde V_2^T};
\end{equation}
accordingly,
\begin{equation}
  Y \approx Q\tilde V_2\tilde \Lambda_2 {\tilde V_2^T}{Q^T}.
\label{eq:equation5}
\end{equation}
Since $QV_2$ is
 orthonormal as a product of orthonormal matrices and $\Lambda_2$ is diagonal,
 this is actually the thin eigenvalue decomposition of $Y$ as in
  Equation (\ref{eq:thin_Y}), obtained without explicitly computing $Y$.
Finally, the solution is given by
\begin{equation}
 Z = Q\tilde V_2{\tilde \Lambda_2 ^{\frac{1}{2}}}.
\label{eq:equation2}
\end{equation}
We call the acceleration involving the decomposition in Section \ref{sec:FMDS_decomposition} \textit{Fast-MDS} (FMDS) and the acceleration involving the decomposition in Section \ref{sec:NMDS_decomposition} \textit{Nystr{\"o}m-MDS} (NMDS). We summarize them in Procedures \ref{alg:FMDS} and \ref{alg:NMDS}.

FMDS and NMDS can be seen as an axiomatic and data-driven approaches that attempt to approximate geodesic distances as a low-rank matrix.
NMDS learns the interpolation operator from the data and obtains better accuracies and time complexities in most of our experiments. In FMDS, the interpolation operator is based on a smoothness assumption through a predefined operator. 
In the FMDS solution, unlike NMDS, the columns of the pairwise distance matrix are interpolated independently.
As a general guideline, when applying the proposed procedures for manifold embedding and pairwise distance approximation, we suggest using NMDS since it is simpler to implement and usually more efficient and accurate.
When dealing with tasks for which boundary conditions vary for different geodesic paths, or where the columns of the distance matrix vary in their structure, such as in matrix completion, FMDS could be used to interpolate the missing values of the columns independently. 
In terms of efficiency, FMDS would be faster if the LBO on the manifold was already computed. Finally, when accuracy is more important for local distances than for larger distances, FMDS may be the preferred choice (see Figures \ref{fig:geodesic_reconstruction_error} and \ref{fig:tri}).

\begin{algorithm}
\floatname{algorithm}{Procedure}
\caption{FMDS}
\begin{algorithmic}[1]

\renewcommand{\algorithmicrequire}{\textbf{Input}}
\renewcommand{\algorithmicensure}{\textbf{Output}}
\Require A manifold $\MM$ represented by $p$ vertices, the number of samples $n$
 and the embedding dimension $m$.
\Ensure A matrix $Z$ which contains the coordinates of the embedding.
\State Choose $n$ vertices from $\MM$ and construct the matrix $R$,
using farthest point sampling described in Section \ref{sec:rows_selection}.
\State Compute the discretized Laplace-Beltrami matrix $L$ of $\MM$ using, for example, cotangent weights \cite{pinkall1993computing}.
\State Compute $M$ according to Equation (\ref{eq:M}).
\State Define $T$, $S$ according Equation \ref{eq:equation3}, and $J$ according to Section \ref{sec:review_MDS}.
\State Compute the QR factorization $Q\mathcal R = JS$.
\State Compute $\tilde V_2$ and $\tilde \Lambda_2$, which contain the $m$ largest eigenvalues
and corresponding eigenvectors of $-\frac{1}{2}\mathcal RT{\mathcal R^T}$, using eigenvalue decomposition.
\State Return the coordinates matrix $Z = Q\tilde V_2 \tilde \Lambda_2^{\frac{1}{2}}$.
\end{algorithmic}
\label{alg:FMDS}
\end{algorithm}

\begin{algorithm}
\floatname{algorithm}{Procedure}
\caption{NMDS}
\begin{algorithmic}[1]

\renewcommand{\algorithmicrequire}{\textbf{Input}}
\renewcommand{\algorithmicensure}{\textbf{Output}}
\Require  A manifold $\MM$ represented by $p$ vertices, the number of samples $n$
 and the embedding dimension $m$.
\Ensure A matrix $Z$ which contains the coordinates of the embedding.

\State Choose $n$ vertices from $\MM$ and construct the matrix $R$,
using farthest point sampling described in Section \ref{sec:rows_selection}.
\State Denote by $R_s$ the rows of $R$ which corresponds to the selected vertices (See Figure \ref{fig:matrixPartition1}).
\State Compute $T = \tilde V \tilde \Lambda^{-1} \tilde V^T$, where $ \tilde \Lambda$ and $\tilde V$ hold the \text{$n_1=\left\lceil\frac{1}{2}n\right\rceil$} largest magnitude eigenvalues and corresponding eigenvectors of $R_s$, and denote $S = R$.
\State Compute the QR factorization $Q\mathcal R = JS$.
\State Compute $\tilde V_2$ and $\tilde \Lambda_2$, which contain the $m$ largest eigenvalues
and corresponding eigenvectors of $-\frac{1}{2}\mathcal RT{\mathcal R^T}$, using eigenvalue decomposition.
\State Return the coordinates matrix $Z = Q\tilde V_2 \tilde \Lambda_2^{\frac{1}{2}}$
\end{algorithmic}
\label{alg:NMDS}
\end{algorithm}

\section{Experimental Results}
\label{sec:Experimental_Results}

\begin{figure}[htbp]
\begin{center}
\includegraphics[width=1\columnwidth]{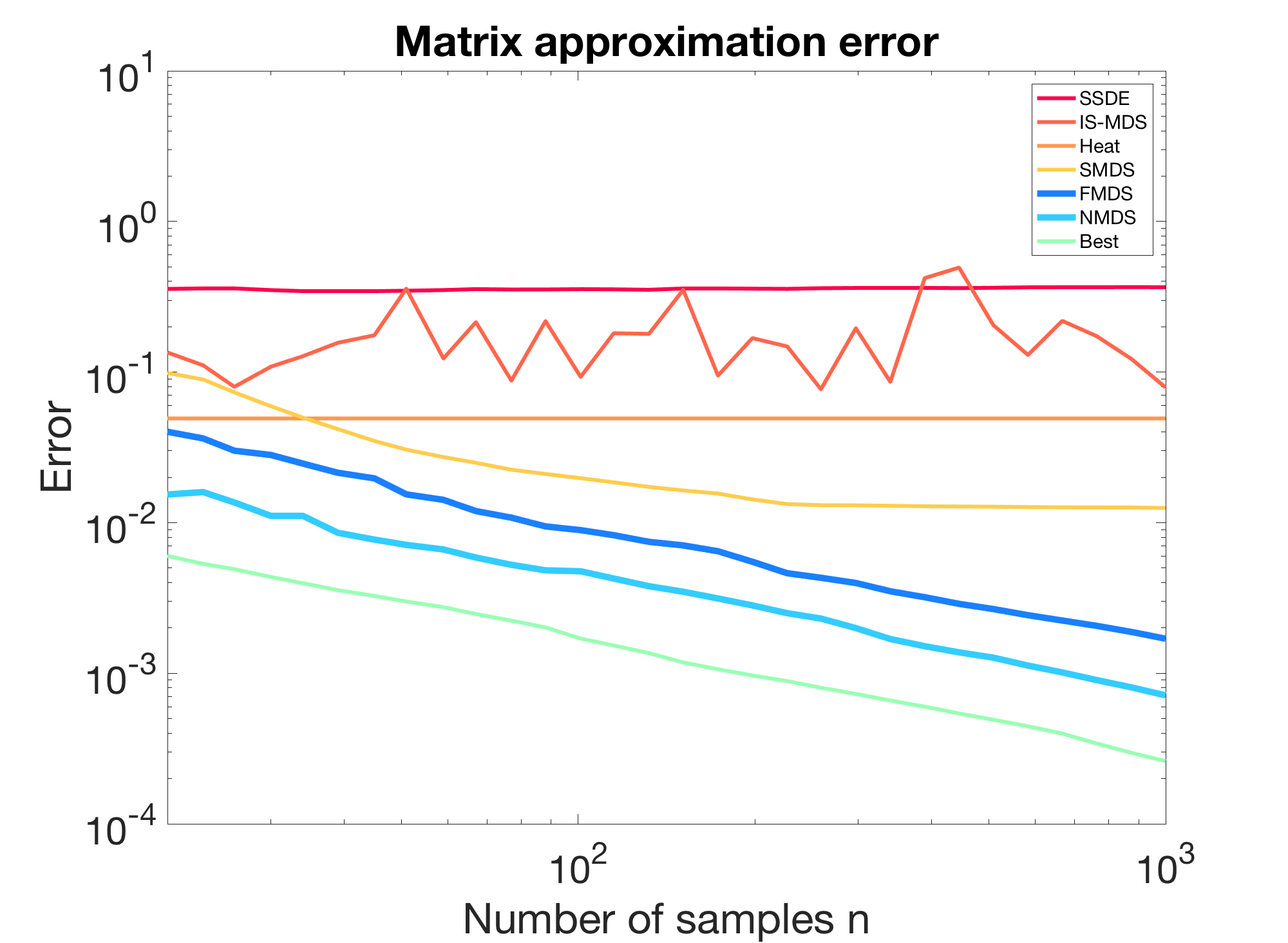}
\end{center}
\caption{The reconstruction error of $D$ with respect to the number of samples $n$, using different methods.}
\label{fig:matrix_reconstruction_error}
\end{figure}

\begin{figure}[htbp]
\begin{center}
\includegraphics[width=0.9\columnwidth]{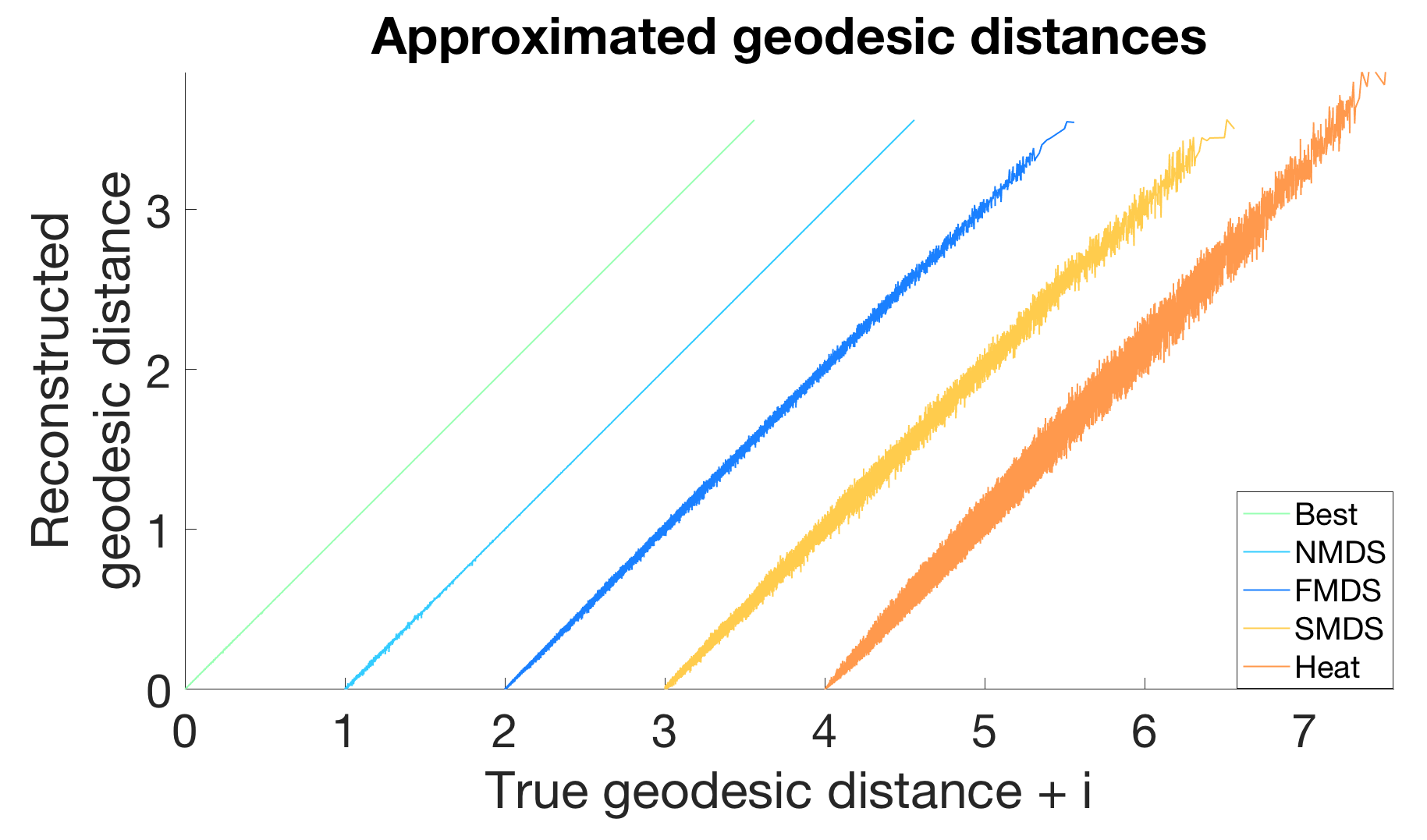}
\end{center}
\caption{Approximated distance as a function of the true distance, using SMDS, FMDS, NMDS, Geodesics in Heat and the best low-rank reconstruction. 
For better visualization, the plots are shifted horizontally by $i=\{0,1,2,3\}$, respectively. 
 In this experiment we used $n=100$.}
\label{fig:geodesic_reconstruction_error}
\end{figure}

Throughout this section, we refer to the classical scaling procedure (Section \ref{sec:review_MDS}) as MDS and compare it to its following approximations:
The proposed Fast-MDS (FMDS, Procedure \ref{alg:FMDS}),
the proposed Nystr{\"o}m-MDS (NMDS), (Procedure \ref{alg:NMDS}),
Spectral-MDS (SMDS \cite{aflalo2013spectral}),
Landmark-Isomap \cite{silva2002global},
Sampled Spectral Distance Embedding (SSDE, \cite{civril2007ssde}), ISOMAP using Nystr{\"o}m and incremental sampling 
  (IS-MDS \cite{yu2012isomap}).
We also compare our distance approximation to the \textit{Geodesics in Heat} (Heat) method \cite{crane2013geodesics}, which computes geodesic distances via a pointwise transformation of the heat kernel, 
 and to the \textit{Constant-time all-pairs distance query} (CTP) method \cite{xin2012constant}, which approximates any pair of geodesic distances
 in a constant time, at the expense of a high complexity pre-processing step.
In Figures \ref{fig:matrix_reconstruction_error} and \ref{fig:geodesic_reconstruction_error}, by `Best' we refer to the best rank-$n$ approximation of a matrix with respect to the 
 Frobenious norm. 
For symmetric matrices this is known to be given by the thin eigenvalue decomposition, keeping only the $n$ eigenvalues with the largest magnitude and $n$ corresponding eigenvectors.
Unless specified otherwise, we use the non-rigid shapes from the SHREC2015 database \cite{3dor.20151064} for the experiments, and each shape is down-sampled to approximately $4000$ vertices. For SMDS we use $100$ eigenvectors. The parameter $\mu$ is set to $10^4$.

\begin{figure}[htbp]
\begin{center}
\begin{tabular}{c c c c c c}
{Giraffe} &
{\includegraphics[width=0.09\columnwidth, height=0.15\columnwidth]{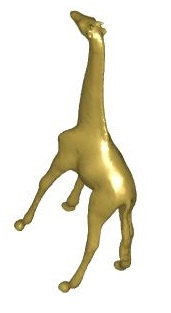} } &
{\includegraphics[width=0.09\columnwidth, height=0.15\columnwidth]{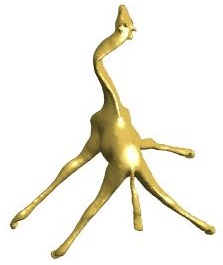} } &
{\includegraphics[width=0.09\columnwidth, height=0.15\columnwidth]{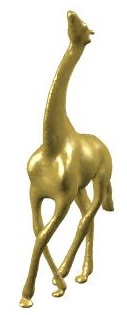} } &
{\includegraphics[width=0.09\columnwidth, height=0.15\columnwidth]{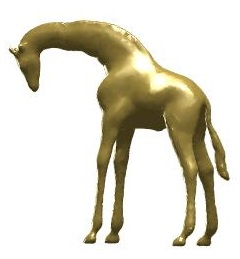} } &
{\includegraphics[width=0.09\columnwidth, height=0.15\columnwidth]{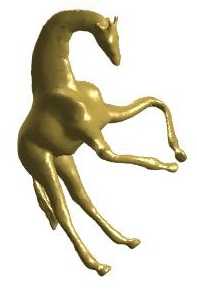} }\\
\hline
{Canonical form} &
{\includegraphics[width=0.09\columnwidth, height=0.15\columnwidth]{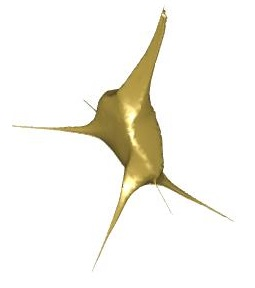} } &
{\includegraphics[width=0.09\columnwidth, height=0.15\columnwidth]{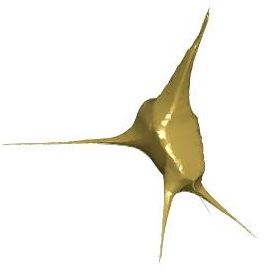} } &
{\includegraphics[width=0.09\columnwidth, height=0.15\columnwidth]{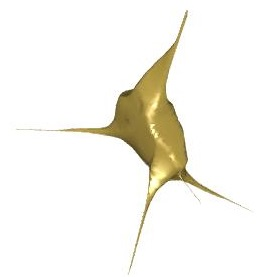} } &
{\includegraphics[width=0.09\columnwidth, height=0.15\columnwidth]{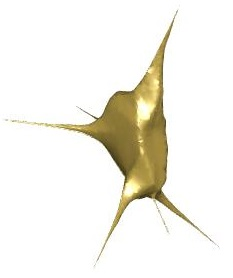} } &
{\includegraphics[width=0.09\columnwidth, height=0.15\columnwidth]{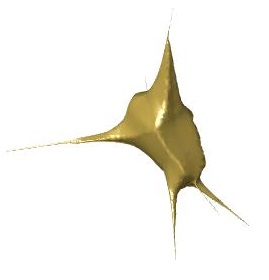} } \\
\hline\hline
{Paper} &
{\includegraphics[width=0.09\columnwidth, height=0.15\columnwidth]{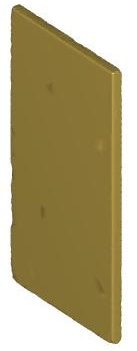} } &
{\includegraphics[width=0.09\columnwidth, height=0.15\columnwidth]{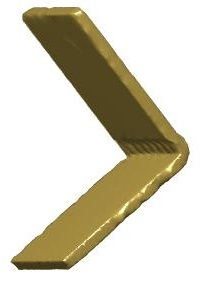} } &
{\includegraphics[width=0.09\columnwidth, height=0.15\columnwidth]{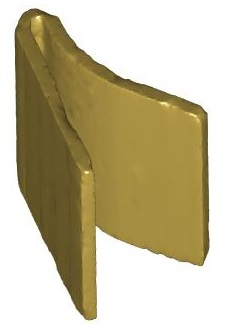} } &
{\includegraphics[width=0.09\columnwidth, height=0.15\columnwidth]{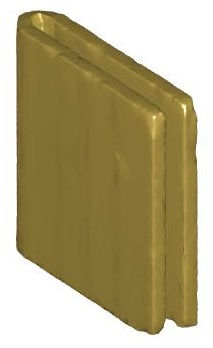} } &
{\includegraphics[width=0.09\columnwidth, height=0.15\columnwidth]{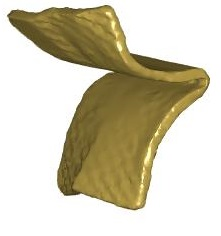} }\\
\hline
{Canonical form} &
{\includegraphics[width=0.09\columnwidth, height=0.15\columnwidth]{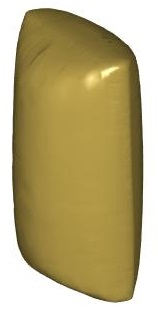} } &
{\includegraphics[width=0.09\columnwidth, height=0.15\columnwidth]{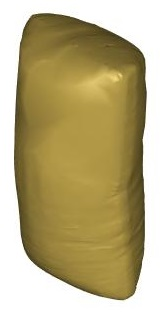} } &
{\includegraphics[width=0.09\columnwidth, height=0.15\columnwidth]{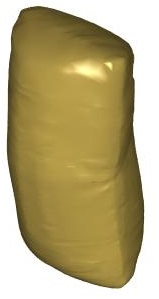} } &
{\includegraphics[width=0.09\columnwidth, height=0.15\columnwidth]{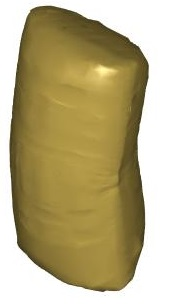} } &
{\includegraphics[width=0.09\columnwidth, height=0.15\columnwidth]{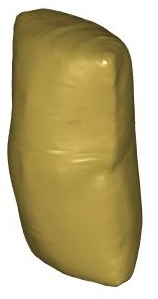} }\\
\hline
\hline
{Watch} &
{\includegraphics[width=0.09\columnwidth, height=0.15\columnwidth]{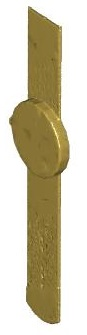} } &
{\includegraphics[width=0.09\columnwidth, height=0.15\columnwidth]{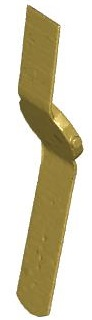} } &
{\includegraphics[width=0.09\columnwidth, height=0.15\columnwidth]{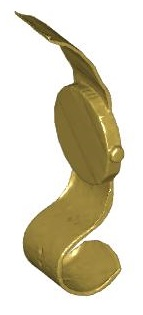} } &
{\includegraphics[width=0.09\columnwidth, height=0.15\columnwidth]{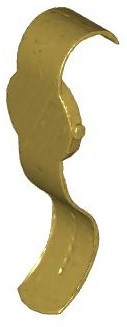} } &
{\includegraphics[width=0.09\columnwidth, height=0.15\columnwidth]{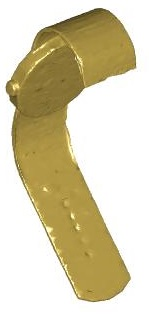} }\\
\hline
{Canonical form} &
{\includegraphics[width=0.09\columnwidth, height=0.15\columnwidth]{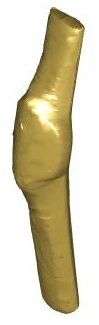} } &
{\includegraphics[width=0.09\columnwidth, height=0.15\columnwidth]{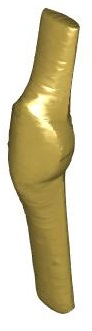} } &
{\includegraphics[width=0.09\columnwidth, height=0.15\columnwidth]{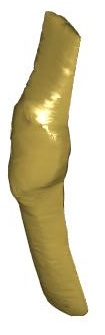} } &
{\includegraphics[width=0.09\columnwidth, height=0.15\columnwidth]{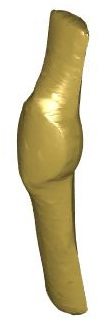} } &
{\includegraphics[width=0.09\columnwidth, height=0.15\columnwidth]{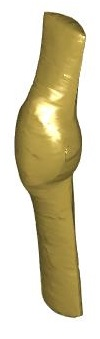} }\\
\end{tabular}
\end{center}
\caption{Shapes in different poses from the SHREC database, and their corresponding canonical forms obtained by NMDS. Here, each shape consists of $10^4$ vertices.}
\label{fig:canonicalForms}
\end{figure}

The output of MDS, known as a \textit{canonical form}
 \cite{elad2003bending}, is invariant to isometric deformations
 of its input. 
We demonstrate this idea in Figure \ref{fig:canonicalForms}, showing
 nearly isometric deformations of the giraffe, paper and watch shapes 
 and their corresponding canonical forms in $\RR^3$ obtained using 
 the NMDS method described in Procedure \ref{alg:NMDS}.
It can be seen that the obtained canonical forms are approximately
 invariant to the different poses.

\begin{figure}[htbp]
\begin{center}
\begin{tabular}{|c|c|c|c|c|c|c|}

\hline
   & {\footnotesize MDS} & {\footnotesize NMDS} & {\footnotesize FMDS} & {\footnotesize SMDS} & {\footnotesize Landmark} & {\footnotesize SSDE}\\
  \hline

{\includegraphics[width=0.07\columnwidth]{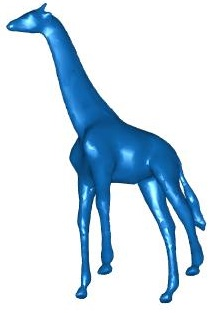} } &
{\includegraphics[width=0.07\columnwidth]{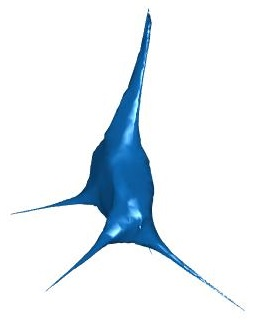} } & 
{\includegraphics[width=0.07\columnwidth]{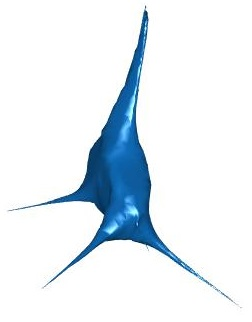} } & 
{\includegraphics[width=0.07\columnwidth]{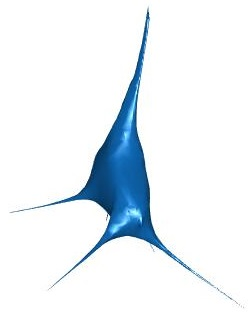} } & 
{\includegraphics[width=0.065\columnwidth]{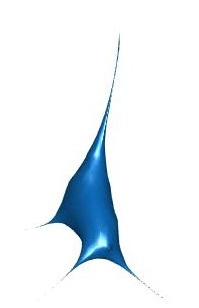} } & 
{\includegraphics[width=0.07\columnwidth]{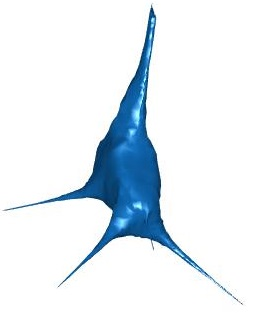} } & 
{\includegraphics[width=0.07\columnwidth]{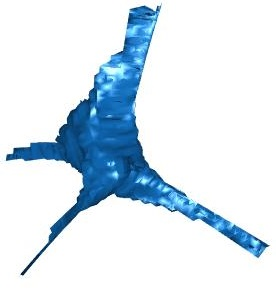} }\\
\hline
\textbf{Stress} & {\scriptsize 5.355} & {\scriptsize 5.355} & {\scriptsize 5.366} & {\scriptsize 5.557} & {\scriptsize 6.288} & {\scriptsize 10.748} \\
\hline

{\includegraphics[width=0.07\columnwidth]{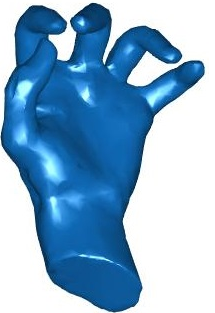} } &
{\includegraphics[width=0.07\columnwidth]{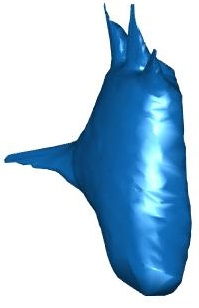} } & 
{\includegraphics[width=0.07\columnwidth]{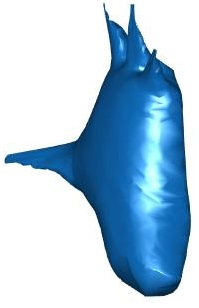} } & 
{\includegraphics[width=0.07\columnwidth]{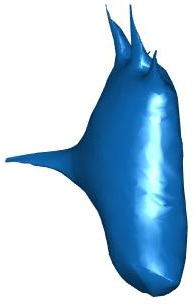} } & 
{\includegraphics[width=0.07\columnwidth]{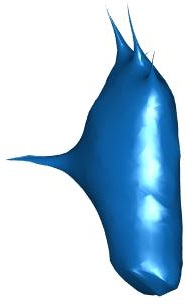} } &
{\includegraphics[width=0.07\columnwidth]{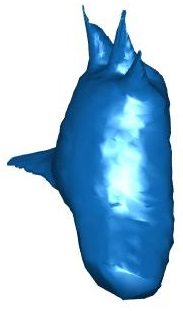} } & 
{\includegraphics[width=0.07\columnwidth]{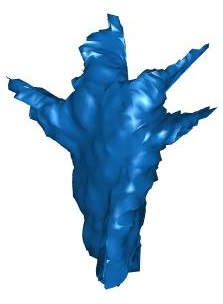} }\\
\hline
\textbf{Stress} & {\scriptsize 3.883} & {\scriptsize 3.887} & {\scriptsize 3.936} & {\scriptsize 4.472} & {\scriptsize 9.128} & {\scriptsize 11.191}\\
\hline
\end{tabular}
\end{center}
\caption{Canonical forms of giraffe and hand shapes, using $n=100$ samples for the compared methods. The stress error $\frac{100}{p^2}{\left\| {Z{Z^T} + \frac{1}{2}JEJ} \right\|_F}$ of each embedding is displayed at the bottom of its corresponding form.}
\label{fig:canonicCompare}
\end{figure}

Unless a surface is an isometric deformation of another flat surface, 
 flattening it into a canonical form would unavoidably involve 
 some deformation of its intrinsic geometry. 
This deformation is termed the \textit{embedding error}, which is 
 defined by the value of the objective function of MDS,
\begin{equation}
\mbox{stress}(Z) = \left \| ZZ^T + \frac{1}{2}JEJ \right \|_F.
\label{eq:stress}
\end{equation}
In our next experiment, shown in Figure \ref{fig:canonicCompare}, we compare the canonical forms of the giraffe and hand shapes, obtained using the different methods. For presentation purposes we scale each embedding
 error to  $\frac{100}{p^2}\mbox{stress}(Z)$ and show it below its
 corresponding canonical form, where $p$ is the number of vertices. 
Qualitative and quantitative comparisons show that the proposed methods 
 are similar to MDS in both minimizers and minima of the objective
 function. 
Additionally, the embedding results of NMDS are similar to those
 of MDS up to negligible differences.

\begin{figure}[htbp]
\begin{center}
\includegraphics[width=1\columnwidth]{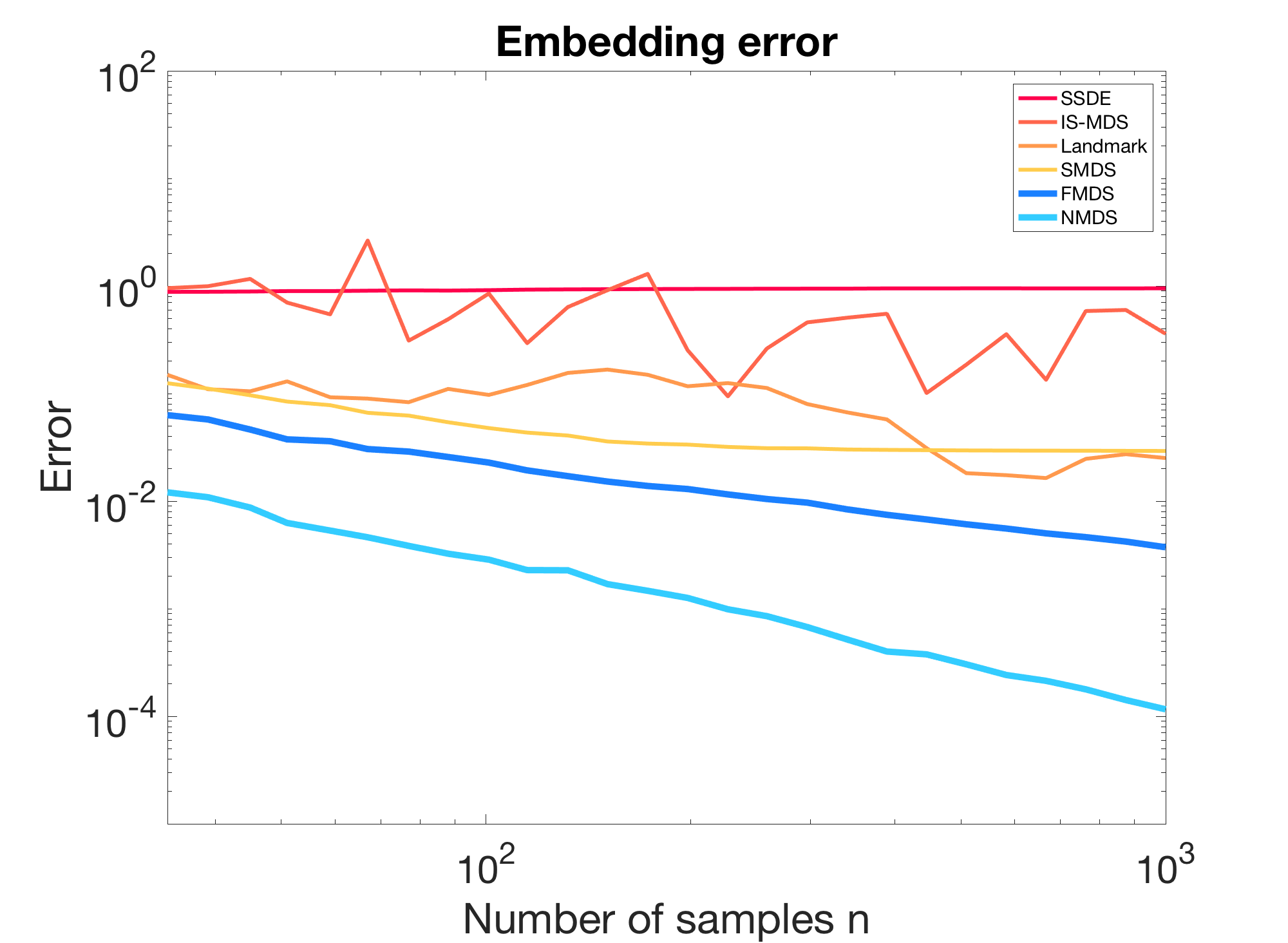}
\end{center}
\caption{Distance of approximated canonical forms to the true one with respect the number of samples $n$, using different methods.}
\label{fig:stress}
\end{figure}

In our next experiment, we directly compare the canonical forms of the different methods ($Z$) to that of MDS ($Z^*$). 
To compare two  
 canonical forms, we align them using ICP \cite{besl1992method} and then 
 compute the relative approximation error
 $\frac{\|Z-Z^*\|_F}{\|Z^*\|_F}$.
Figure \ref{fig:stress} shows the results with respect to the number
 of samples $n$, on the hand shape.
It can be seen that both proposed methods outperform the others
 in approximating the embedding of MDS.

The proposed methods not only provide fast and accurate approximations compared to MDS, but can also be used to significantly accelerate the computation of geodesic distances by approximating them. In the following experiments, we measure the distance approximation error on the giraffe shape.
In Figure \ref{fig:matrix_reconstruction_error}, we compare the relative reconstruction error 
 $\frac{\|\hat D-D\|_F}{\|D\|_F}$, where $\hat D$ approximates $D$ using different methods.
Since the compared methods are essentially different low-rank approximations of $E$, we also add the best rank-$n$ approximation to the comparison.
 
In Figure \ref{fig:geodesic_reconstruction_error}, we visualize the geodesics approximations of randomly chosen $30K$ pairs of distances using different methods. We do this by plotting the true and approximated distances along the $x$ and $y$ axes, thus showing the approximation's distribution over the surface. 

\begin{figure}[htbp]
\begin{center}
\includegraphics[width=1\columnwidth]{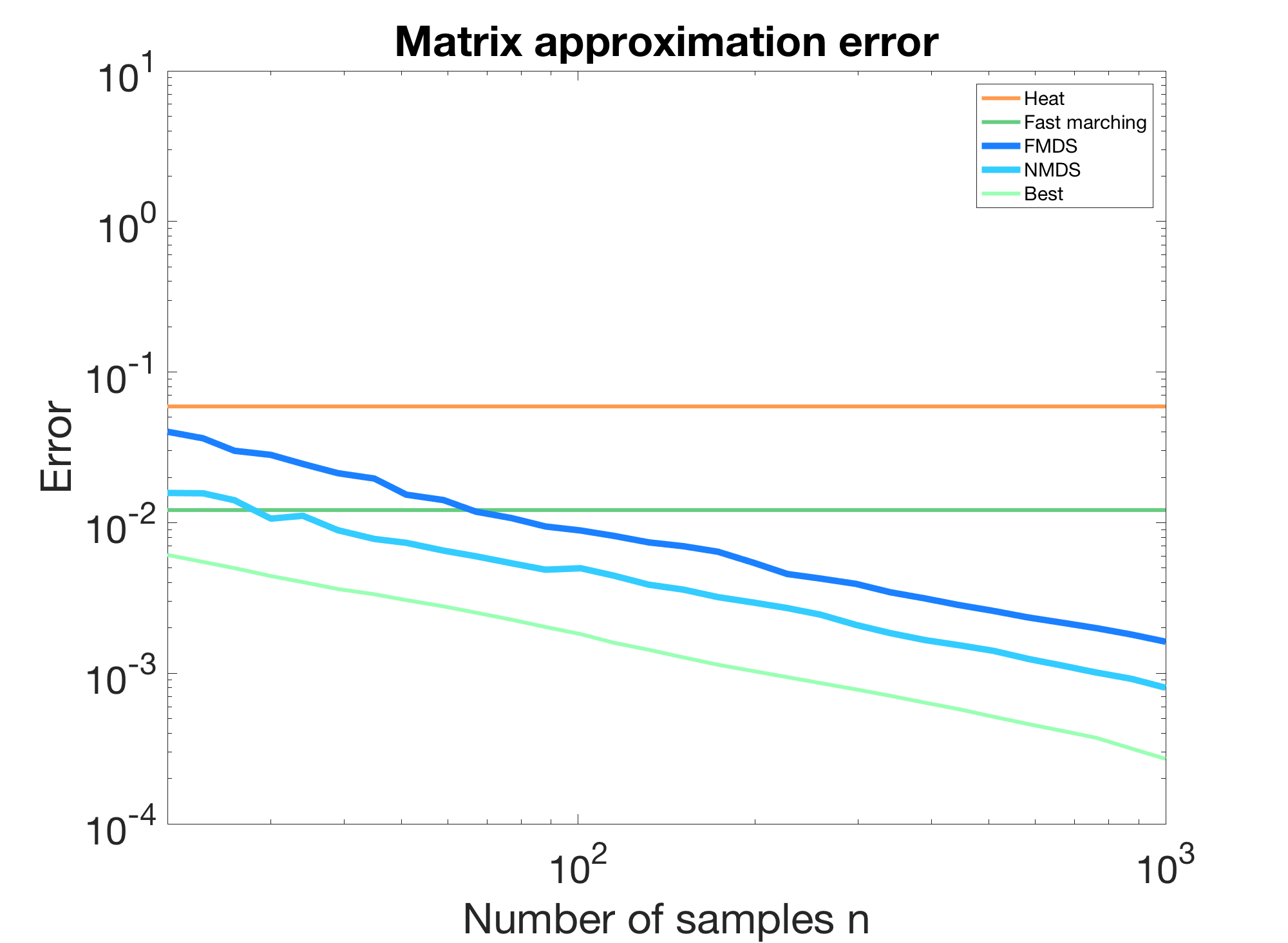}
\end{center}
\caption{The reconstruction error of $D$ with respect to the number of samples $n$ as evaluated by exact geodesics, using different methods.}
\label{fig:results_giraffe_exact}
\end{figure}

Thus far, we used fast marching as a gold standard in order to compute the initial set of geodesic distances and then evaluate the results. 
In the following experiment, we use the exact geodesics method \cite{surazhsky2005fast} instead. This method computes the exact geodesic paths on triangulated surfaces and is more accurate in terms of truncation error when compared to the limiting continuous solution. 
This should also allow us to compare our derived geodesic distances to the ones obtained using fast marching. 
Figure \ref{fig:results_giraffe_exact} shows that exact geodesic distances from less than $30$ points were required in the initialization step,
in order to obtain a full matrix of pairwise geodesic distances, computed using NMDS, which is more accurate than the full matrix obtained by the 
fast marching method.
This experiment demonstrates the average error.

In Table \ref{fig:constant_time} we compare the preprocessing, query times and accuracies of CTP to the one proposed in this paper. 
We use the same Root-Mean-Square, over a set of random pairs, of the relative error 
$\epsilon(i,j)= \frac{\hat D(i,j)- D(i,j)}{D(i,j)}$, which was used in their paper. 
In this experiment, 
the initial set of distances was computed using fast marching for NMDS and FMDS, and exact geodesics for CTP. 
All three methods were evaluated using exact geodesics as the ground truth.
The query time complexity is $O(1)$ for CTP and $O(n)$ for FMDS and NMDS, where $n = 20$--$100$ is the number of samples. 
Nevertheless, in the method we propose, query operations are done by one matrix product that typically has efficient implementations in standard frameworks, and thus our query step is more than $\times 1000$ faster.
As for the pre-processing step, the method we introduce requires many fewer samples (around $\times 100$ less geodesics computations) and is thus also much faster.

\begin{table}[htbp]
\begin{center}
\includegraphics[width=1\columnwidth]{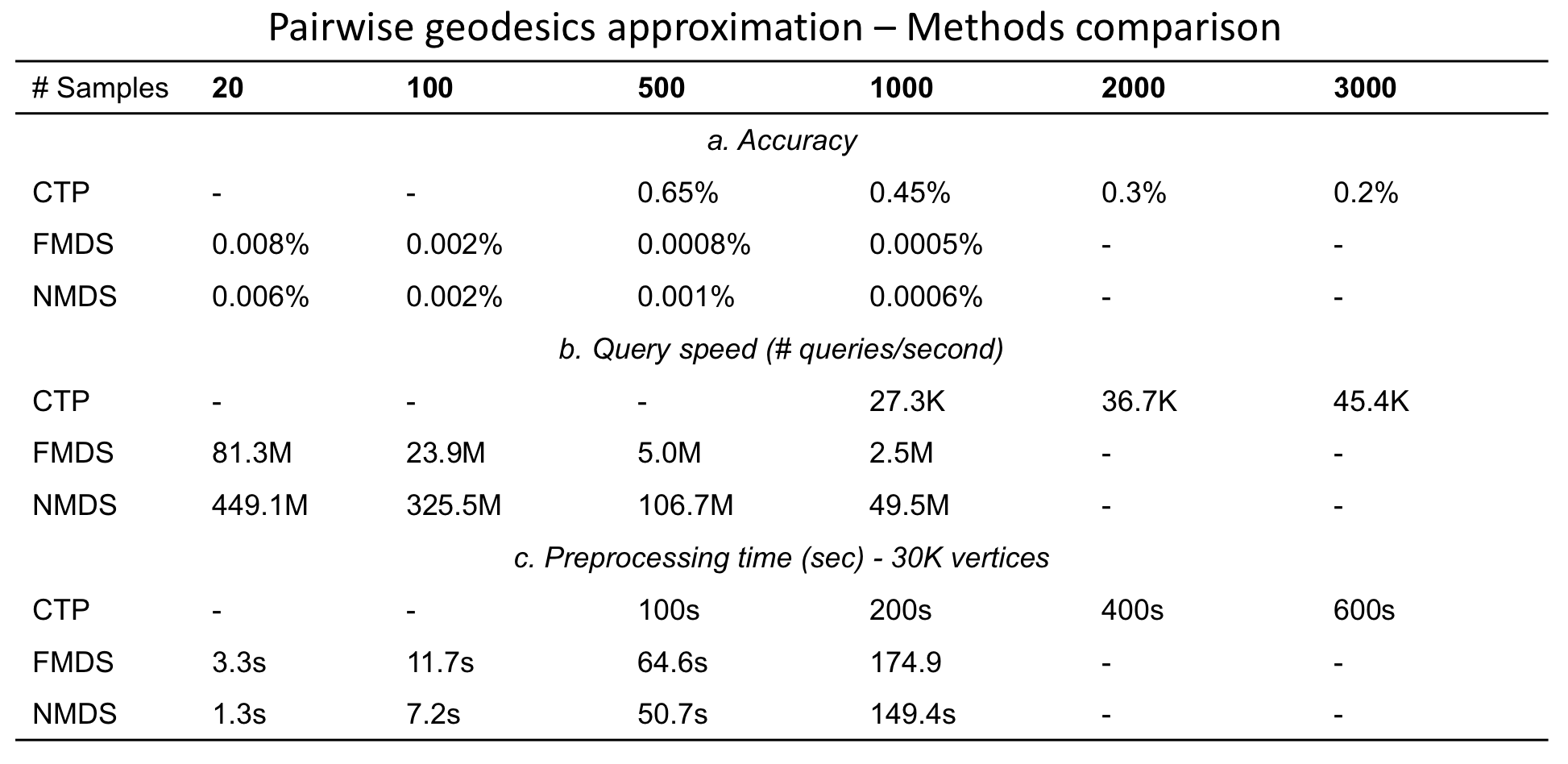}
\end{center}
\caption{(a) Accuracy, (b) query, and (c) preprocessing time of geodesic distance approximations using FMDS, NMDS and CTP.}
\label{fig:constant_time}
\end{table}

\begin{figure}[htbp]
\centering
\includegraphics[width=0.4\columnwidth]{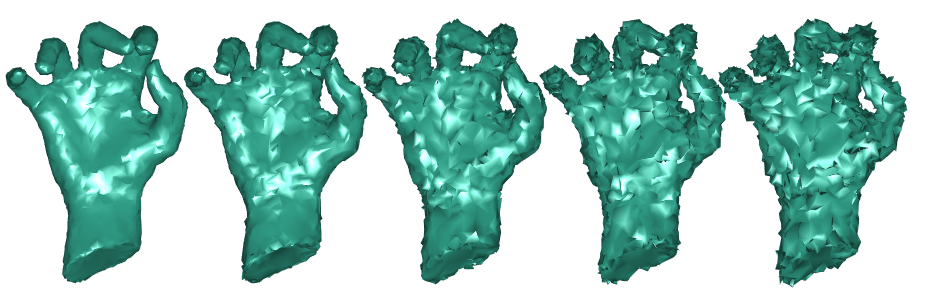} 
\hspace{8mm}
\includegraphics[width=0.4\columnwidth]{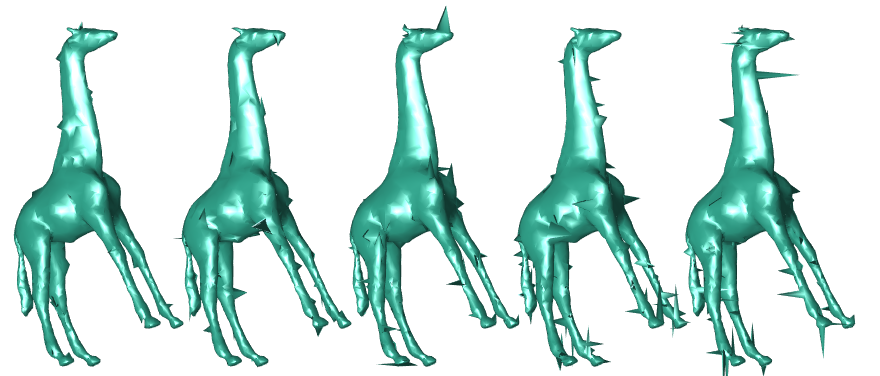}
\includegraphics[width=0.49\columnwidth]{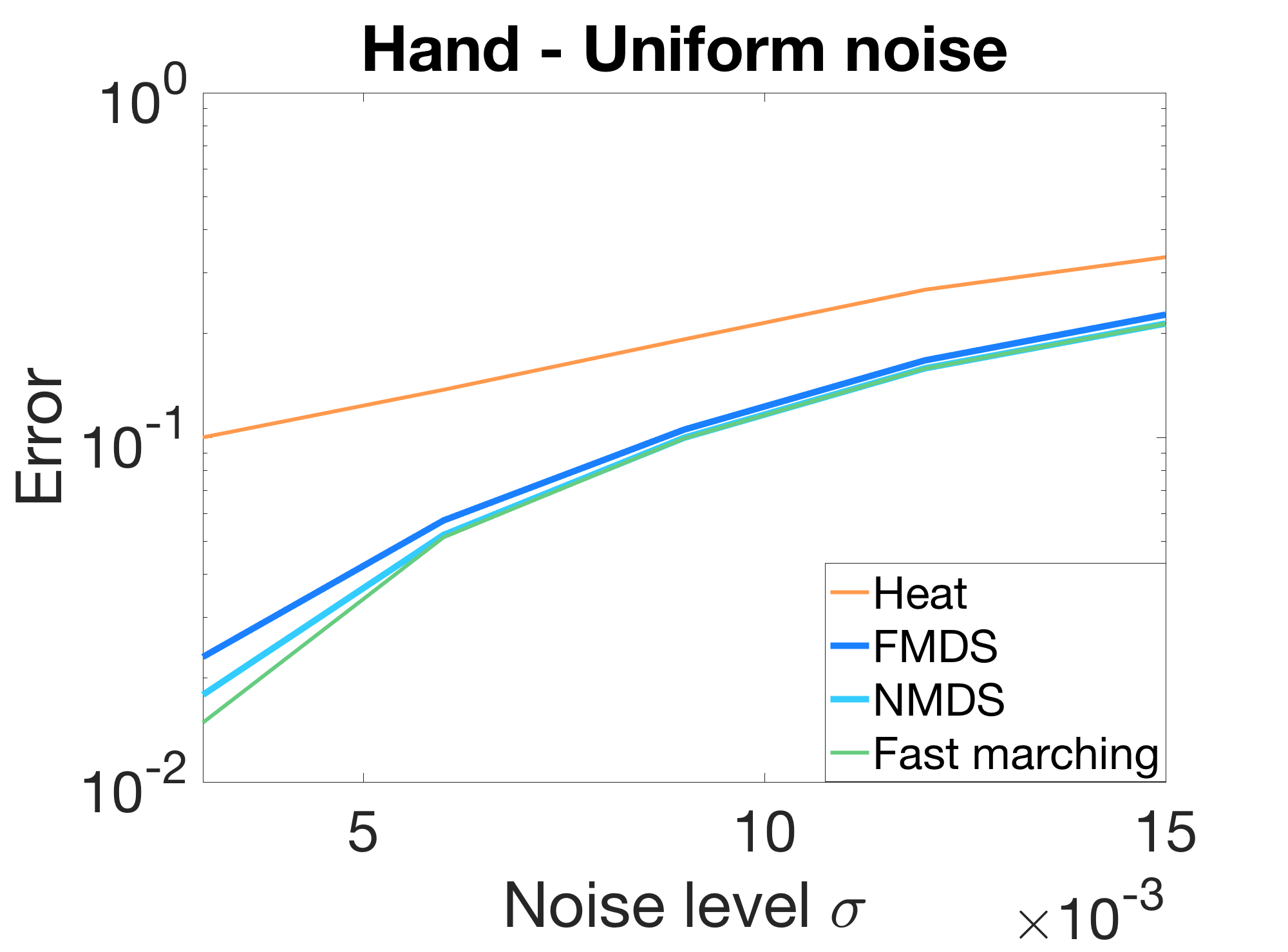}
\includegraphics[width=0.49\columnwidth]{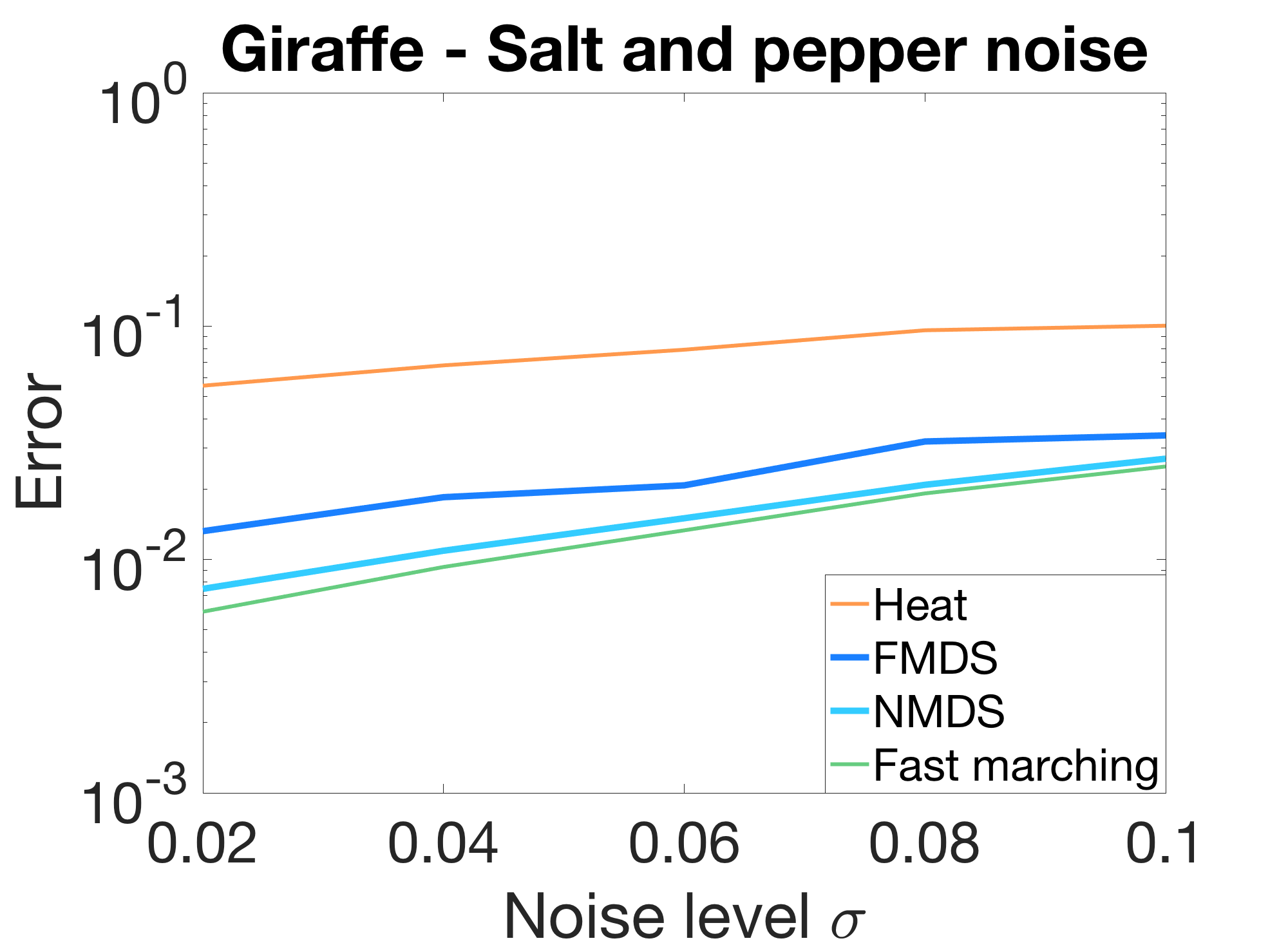}

\caption{Influence of noise on the geodesic approximation error for Geodesics in Heat, FMDS, NMDS, and fast marching. All approximations are compared to the geodesic distances that were computed using fast marching before adding the noise.}
\label{fig:results_noise}
\end{figure}

Next, we evaluate the influence of noise on the geodesic distance approximations.
We add random uniform and sparse (salt and pepper) normally distributed noise to the hand and giraffe shapes, respectively. 
The sparse noise was added to $1\%$ of the vertices.
We initialize NMDS and FMDS with the fast marching distances that were computed on the noisy surfaces, and evaluate the resulting geodesics using the fast marching distances that were computed before adding the noise.
Figure \ref{fig:results_noise} shows that the distances computed by our methods remain close to the noisy fast marching distances. 
Thus, we conclude that the methods we propose approximate the distances they 
 were initialized with well, whether they contain noise or not.

\begin{figure}[htbp]
\begin{center}
\includegraphics[width=0.8\columnwidth]{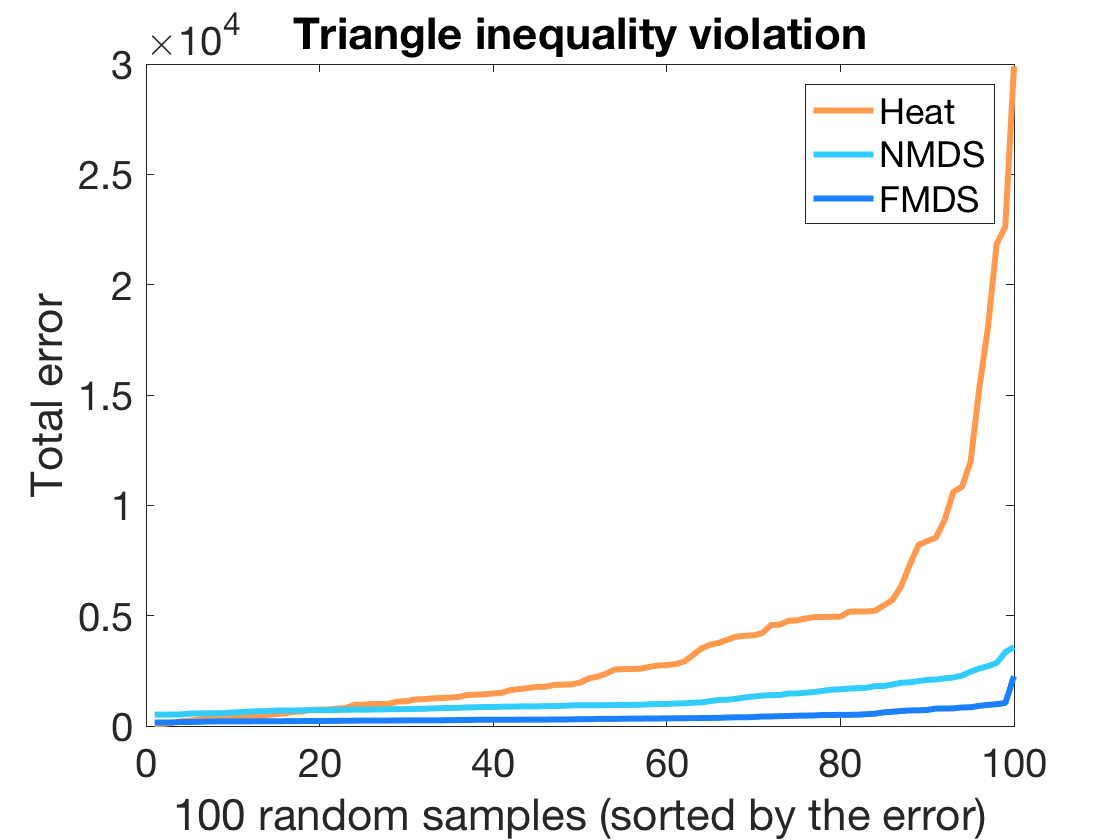}
\end{center}
\caption{
The triangle inequality total violation error of $100$ randomly chosen surface points (sorted), for NMDS, FMDS and Geodesics in Heat.
}
\label{fig:tri}
\end{figure}

Next, we evaluate the deviation of the approximated distance from satisfying the triangle inequality, as done in \cite{crane2013geodesics}.
We compute the total violation error of a surface point $a$ as follows.
For any other two surface points $b,c$, if 
\begin{equation}
{\cal E}rr(a,b,c) = \|b-c\| - (\|b-a\| + \|c-a\|)    
\end{equation}
is positive, we add ${\cal E}rr(a,b,c)$ to the total violation error of point $a$.
We choose $100$ random points on the surface and measure their total violation error. 
Figure \ref{fig:tri} compares the NMDS, FMDS, and Heat methods.
It can be seen that NMDS and FMDS had lower violation errors than Geodesics in Heat. Among the proposed methods, FMDS has the lowest error since it is more accurate than NMDS in approximating the small distances.

\begin{figure}[htbp]
\begin{center}
\includegraphics[width=0.8\columnwidth]{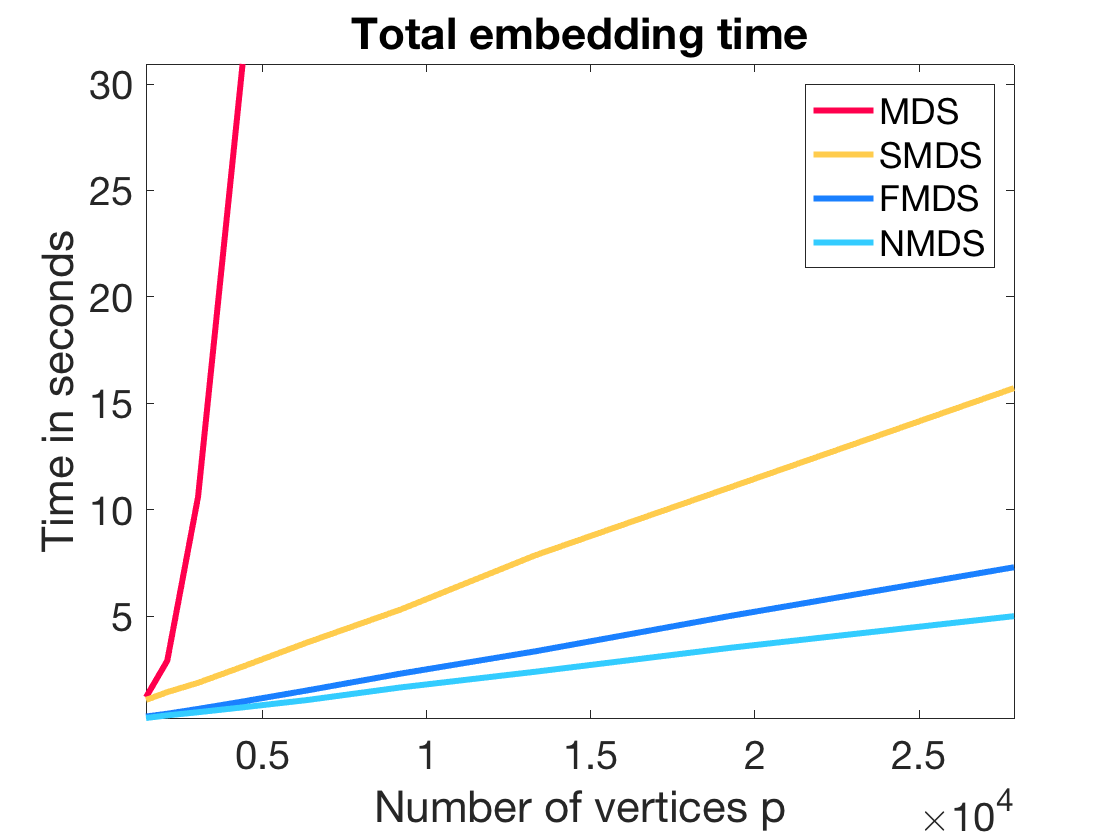}
\end{center}
\caption{Average computation times (in seconds) for MDS, NMDS, FMDS and SMDS with respect to the number of vertices.}
\label{fig:total_time}
\end{figure}

Finally, as shown in Figure \ref{fig:total_time}, we evaluate the average
computation time of MDS, NMDS, FMDS, and SMDS on three shapes from the 
 TOSCA database \cite{bronstein2008numerical}, including the 
 computation of the geodesic distances. 
We change the number of vertices $p$ by subsampling and re-triangulating
 the original shapes.
The computations were evaluated on a 2.8 GHz i7 Intel computer with
 16GB RAM.
Due to time and memory limitations, MDS was computed only for small
 values of $p$.

 \section{Sphere embedding}
 \label{sec:applications}

So far we discussed MDS as a \textit{flattening} procedure, meaning that the data is embedded into a flat Euclidean space. 
Nevertheless, in some cases, it may be beneficial to embed the data onto a sphere, on which distances are measured according to the sphere's metric.
 In this section, we extend the proposed methods for embedding onto a sphere rather than in a Euclidean space. 
 
A $k$-dimensional sphere is the set of all points in $\RR^{k+1}$ at an equal distance from a fixed point.
Without loss of generality, we assume that this fixed point is the origin, so that the sphere is centered.
The geodesic distance between two points $z_i$ and $z_j$ on a sphere of radius $r$ can be computed by the arc length \begin{equation}
\frac{\alpha_{ij}}{2\pi}2\pi r = \alpha_{ij} r,
\end{equation}
where $\alpha_{ij}$ is the angle between the vectors $z_i, z_j$ in $\RR^{k+1}$.
Given a manifold with pairwise geodesic distances $D_{ij}$, we aim to find a set of points $\{z_i\}_{i=1}^p$ in $\RR^{k+1}$, representing the embedding of the manifold on a sphere, such that $D_{ij} \approx \alpha_{ij} r$. 
Dividing by $r$ and applying $\cos$ on both sides leads to $\cos{\frac{D_{ij}}{r}} \approx \cos{\alpha_{ij}}$

Let $D$ be a $p \times p$ matrix holding the distances $D_{ij}$, and define $Z_r=\frac{1}{r^2}Z$.
Since $\cos \alpha_{ij} = \frac{z_i^Tz_j}{|z_i||z_j|}$, we can write in matrix formulation,
\begin{equation}
\cos\left (\frac{D}{r}\right ) \approx Z_rZ_r^T.
\end{equation}
Then, we can formulate the problem through the minimization
\begin{equation}
  Z_r^* = \arg \min {\left\| {Z_rZ_r^T + E} \right\|_F},
\end{equation}
similar to classical scaling, 
where $E = \cos(\frac{D}{r})$. 
This minimization can be solved using the same techniques presented
 in FMDS or NMDS. 
Namely, compute the decomposition of $E$ with one of the methods in Section \ref{sec:Distances_Interpolation}. 
Then, follow Section \ref{sec:acceleration} to obtain the truncated
 eigenvalue decomposition while ignoring the matrix $J$ and the 
  scalar $\frac{1}{2}$.
The final solution is given by $Z=r^2Z_r^*$.

A step of normalization of the rows of $Z_r^*$ can be added to constrain the points to lie on the sphere. 
Notice that when there exists an exact spherical solution, it will be obtained without this step. 
This results from the fact that $D_{ii} = 0$ and, therefore, $E_{ii} = 1$. 
Hence, for an exact solution without embedding errors, we get
 $\{Z_rZ_r^T\}_{ii} = 1$, which is true only when the rows of $Z_r$ 
 are normalized.

In the following experiment, a camera was located on a chair at the 
 center of a room, and a video was taken while the chair was spinning. 
The frames of the video can be thought of as points in a high-dimensional
 space, lying on a manifold with a spherical structured intrinsic geometry. Using the method described in this section, we embed this manifold onto
 a $1$-dimensional sphere using NMDS (without the normalization step), and show the results in Figure \ref{fig:sphere_embedding}. 
 A nice visualization of this experiment appears in the supplementary material.
Both MDS and NMDS result in a similar circular embedding,
 revealing the intrinsic geometry of the manifold. 
The pose of the camera
  for any frame can then be extracted from the embedding, even if the 
  order of the frames in the video is unknown.
This experiment also demonstrates that the proposed methods are capable 
 of dealing with more complex manifolds in high-dimensional spaces
 and with embeddings in metrics other than Euclidean ones.

 \begin{figure}[htbp]
 \centering
\begin{subfigure}[t]{0.2\textwidth}
\begin{center}
\includegraphics[width=\linewidth]{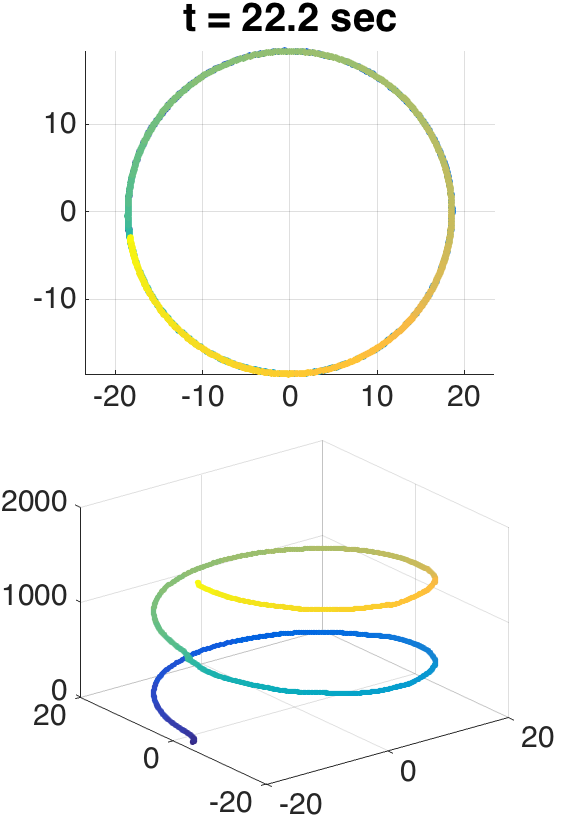}
\end{center}
\caption{\scriptsize MDS}
\label{fig:sphere_embedding_MDS}
\end{subfigure}
\begin{subfigure}[t]{0.2\textwidth}
\begin{center}
\includegraphics[width=\linewidth]{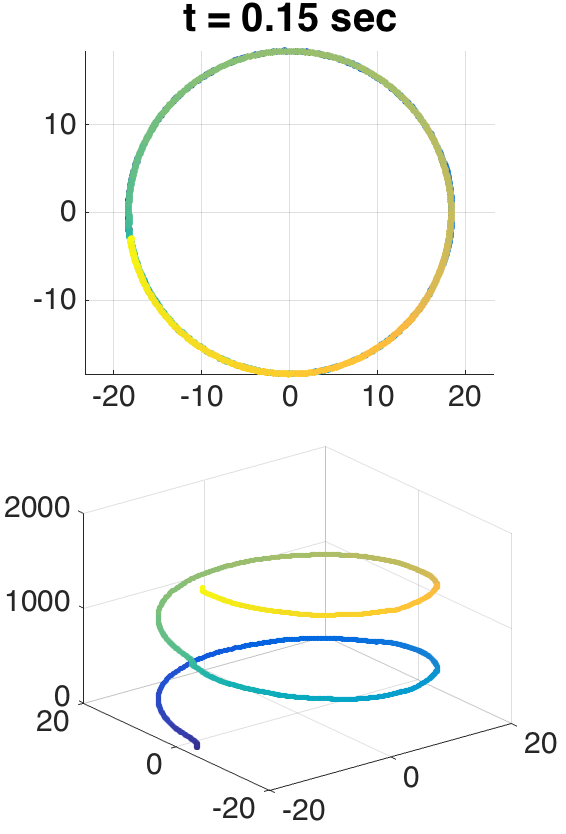}
\end{center}
\caption{\scriptsize NMDS}
\label{fig:sphere_embedding_NMDS}
\end{subfigure}

\caption{Embedding points onto a sphere using Full MDS (left) and the proposed NMDS (right). The points are frames of a video of a room taken from its center while rotating two whole turns. The images at the top show the embedding on a $1$-dimensional sphere. The time it took to create each embedding is added at the top. For better visualization, we add an additional axis of the frame number, shown in the images at the bottom.}
\label{fig:sphere_embedding}
\end{figure}
Next, we randomly sample $p$ points $\{x_i\}_1^p$ from a quarter of a $k$-dimensional sphere of radius $r=1$, centered at the origin.  We build a knn graph from the points such that each point is connected to $w$ neighbors. 
We compute $n$ initial geodesic distances both analytically and using Dijkstra's shortest path for the farthest point sampling step, and approximate the rest of the distances using NMDS. 
Thereafter, we embed the distances on a $\tilde k$-dimensional sphere, as well as in a $\tilde k$-dimensional Euclidean space.
For evaluation, we analytically compute all distances and compare them to the approximated ones using the relative reconstruction error seen previously in Figure \ref{fig:matrix_reconstruction_error}.
The embedding error is evaluated using the stress in Equation \ref{eq:stress}.

\begin{figure}[htbp]
\begin{center}
\includegraphics[width=0.49\columnwidth]{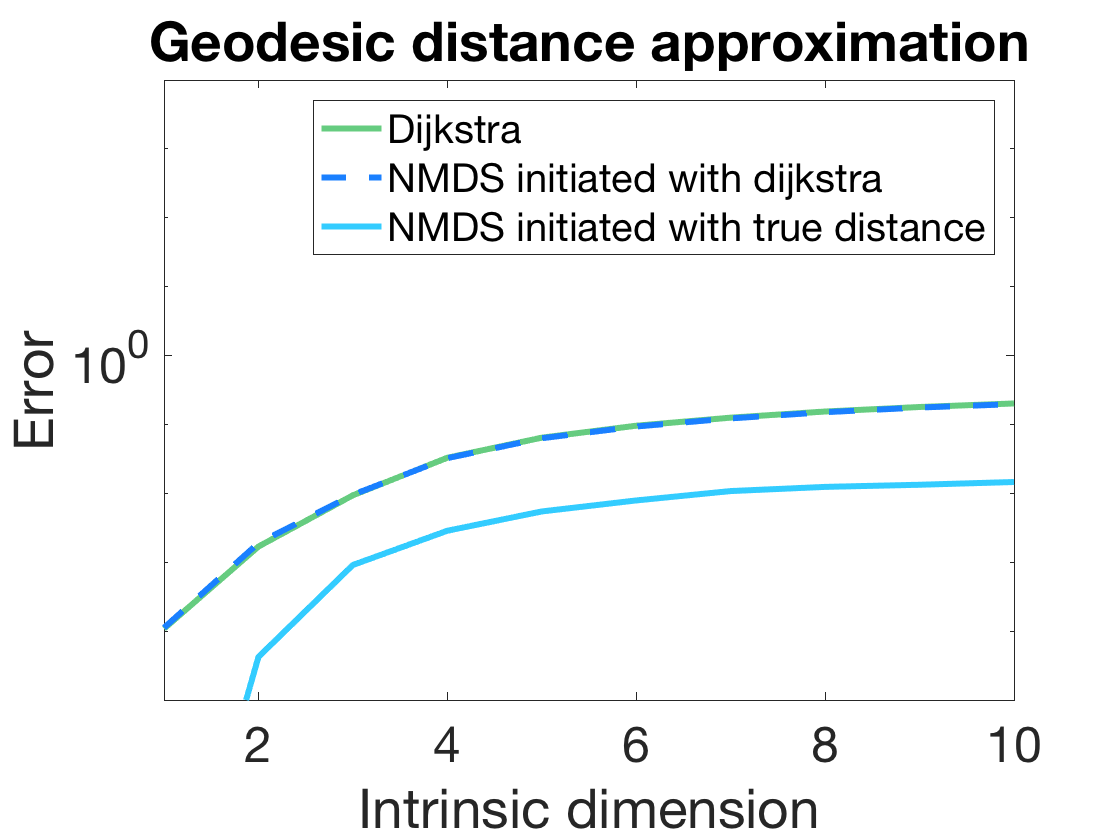}\\
\includegraphics[width=0.49\columnwidth]{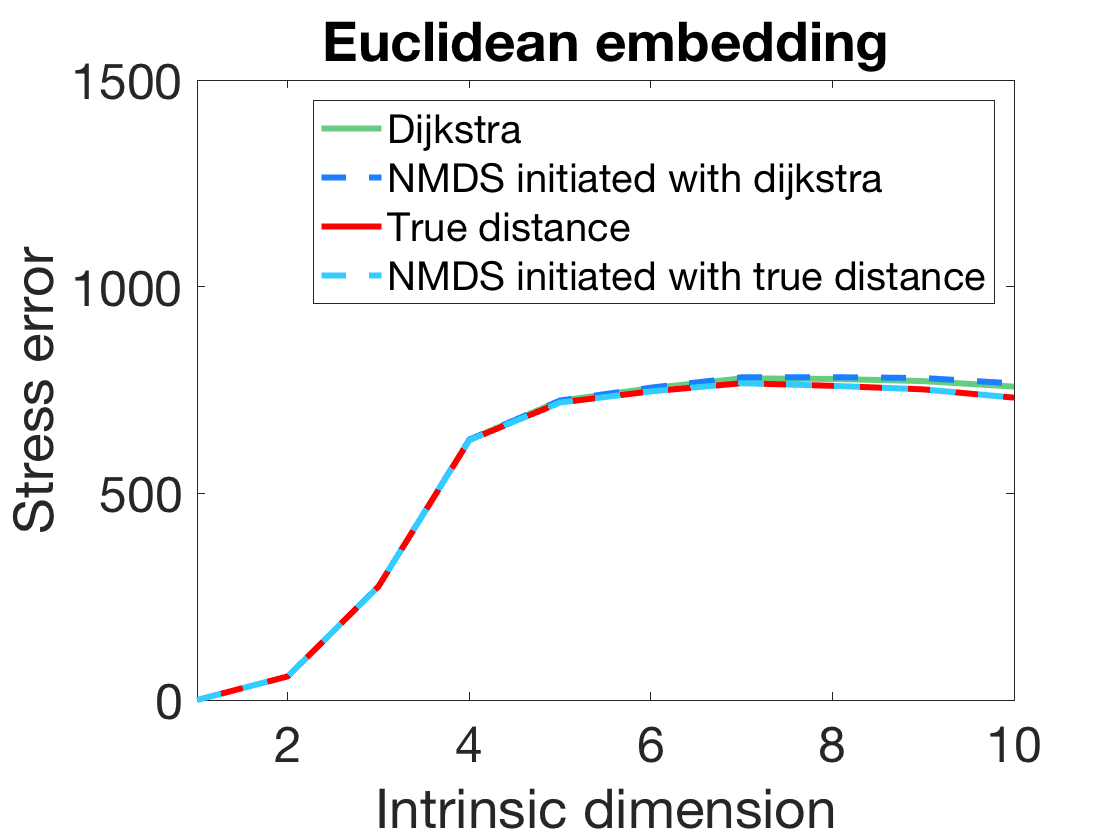}
\includegraphics[width=0.49\columnwidth]{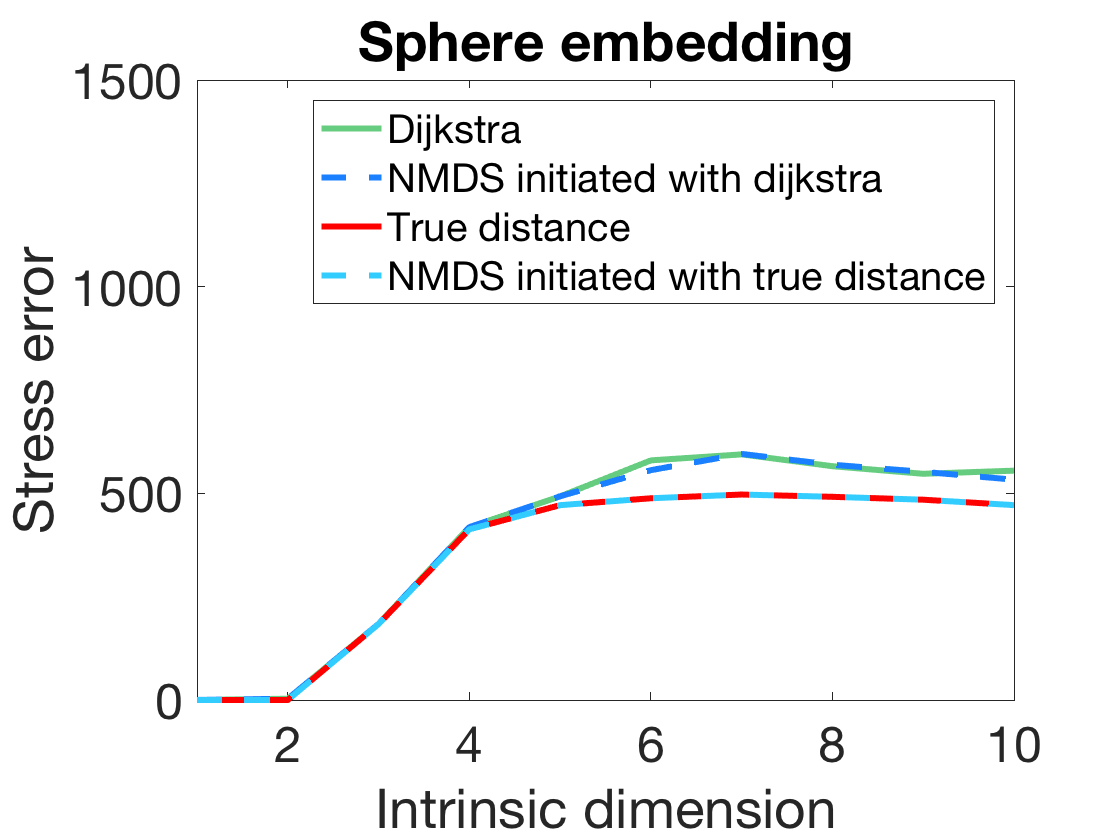}
\end{center}
\caption{
Evaluation of the proposed method with respect to the intrinsic dimension $k$.
Top: The geodesic reconstruction error of Dijkstra's algorithm, NMDS initialized by Dijkstra, and NMDS initialized by analytically computed distances. 
Bottom left: Stress error of embedding in a Euclidean space, using the distances computed analytically (True), with Dijkstra, and with NMDS initialized
once with the analytic and then with the Dijkstra approximation of the distances. 
Bottom right: Stress error when embedding in a low-dimensional sphere.
}
\label{fig:int_dim}
\end{figure}

For this experiment, we choose $p = 2000$ points, $n = 100$ initial distances, embedding space dimension $\tilde k = 2$, and $w = 120$ nearest neighbors. 
Note, however, that the results did not change much for different sets of plausible parameters.
We test the proposed method for different intrinsic dimensions $k$ between $1$ and $10$. 
Figure \ref{fig:int_dim} shows the results. Not surprisingly, the stress is lower for sphere embedding than for Euclidean embedding. 
It can be seen that the proposed method approximates well both the ground truth embedding and the distances it was initialized with, regardless of the intrinsic dimension of the sphere manifold. 
For the distance approximation, the error of NMDS initialized by Dijkstra's algorithm stems mainly from the error introduced by the graph approximating the continuous distance.
As for the stress, since the embedding procedure is robust to noise in distances, the stress is not substantially different between the two proposed methods.

\section{Conclusions}


In this paper we develop two methods for approximating MDS in its classical scaling version. 
In FMDS, following the ideas of Spectral-MDS, we interpolate the geodesic distances from a few samples using a smoothness assumption. This interpolation results in a product of small matrices that is a low-rank approximation of the distance matrix. 
Then, time and space complexities of MDS are significantly reduced by reformulating its minimization through the small matrices. 
The main contribution to the improvement in time complexity and accuracy over SMDS is the fact that we work entirely in the spacial domain, as opposed to SMDS, which translates the problem into the spectral domain and truncates the eigenspace.
FMDS sets the stage for the second method we term NMDS. Instead of constructing an interpolation matrix using smoothness assumptions, it is now learned from the sampled data, which results in a different low-rank approximation of the distance matrix. A small modification of the method shows its relation to the Nystr{\"o}m approximation.
Although NMDS performs better than FMDS in most experiments, the interpolation ideas of FMDS can  also be used for other tasks, such as matrix completion, where instead of being arranged in columns, the known entries might be spread over the matrix.
Experimental results show both methods achieve state-of-the-art results in canonical form computations. 
As a by-product, the methods we introduced can be used to compute geodesic distances on manifolds very fast and accurately.
Finally, an extension to embedding on spheres shows how these methods can deal with more complex manifolds in high-dimensional space embedded into non-flat metric spaces.
\section{Acknowledgements}
The research leading to these results has received funding from the European Research
  Council under European Union's Seventh Framework Programme,
  ERC Grant agreement no. 267414


{\small
\bibliography{refs}

\begin{thebibliography}{10}

\bibitem{aflalo2013spectral}
Y.~Aflalo and R.~Kimmel.
\newblock Spectral multidimensional scaling.
\newblock {\em Proceedings of the National Academy of Sciences},
  110(45):18052--18057, 2013.

\bibitem{arcolano2010Nystrom}
N.~Arcolano and P.~J. Wolfe.
\newblock {Nystr{\"o}m} approximation of wishart matrices.
\newblock In {\em Acoustics Speech and Signal Processing (ICASSP), 2010 IEEE
  International Conference on}, pages 3606--3609. IEEE, 2010.

\bibitem{belkin2001laplacian}
M.~Belkin and P.~Niyogi.
\newblock Laplacian eigenmaps and spectral techniques for embedding and
  clustering.
\newblock In {\em NIPS}, volume~14, pages 585--591, 2001.

\bibitem{besl1992method}
Paul~J Besl and Neil~D McKay.
\newblock Method for registration of 3-d shapes.
\newblock In {\em Robotics-DL tentative}, pages 586--606. International Society
  for Optics and Photonics, 1992.

\bibitem{borg2005modern}
I.~Borg and P.~J. Groenen.
\newblock {\em Modern multidimensional scaling: Theory and applications}.
\newblock Springer, 2005.

\bibitem{botsch2008linear}
Mario Botsch and Olga Sorkine.
\newblock On linear variational surface deformation methods.
\newblock {\em IEEE transactions on visualization and computer graphics},
  14(1):213--230, 2008.

\bibitem{bronstein2008numerical}
A.~M. Bronstein, M.~M. Bronstein, and R.~Kimmel.
\newblock {\em Numerical geometry of non-rigid shapes}.
\newblock Springer, 2008.

\bibitem{campen2011walking}
Marcel Campen and Leif Kobbelt.
\newblock Walking on broken mesh: Defect-tolerant geodesic distances and
  parameterizations.
\newblock In {\em Computer Graphics Forum}, volume~30, pages 623--632. Wiley
  Online Library, 2011.

\bibitem{civril2007ssde}
A.~Civril, M.~Magdon-Ismail, and E.~Bocek-Rivele.
\newblock {SSDE}: Fast graph drawing using sampled spectral distance embedding.
\newblock In {\em Graph Drawing}, pages 30--41. Springer, 2007.

\bibitem{crane2013geodesics}
Keenan Crane, Clarisse Weischedel, and Max Wardetzky.
\newblock Geodesics in heat: A new approach to computing distance based on heat
  flow.
\newblock {\em ACM Transactions on Graphics (TOG)}, 32(5):152, 2013.

\bibitem{drrksrranote}
E.~W. Dijkstra.
\newblock A note on two problems in connexion with graphs.
\newblock {\em Numerische Mathematik l}, pages 269--27.

\bibitem{donoho2003hessian}
D.~L. Donoho and C.~Grimes.
\newblock Hessian eigenmaps: Locally linear embedding techniques for
  high-dimensional data.
\newblock {\em Proceedings of the National Academy of Sciences},
  100(10):5591--5596, 2003.

\bibitem{drineas2006fast}
P.~Drineas, R.~Kannan, and M.~W. Mahoney.
\newblock Fast monte carlo algorithms for matrices iii: Computing a compressed
  approximate matrix decomposition.
\newblock {\em SIAM Journal on Computing}, 36(1):184--206, 2006.

\bibitem{drury1996computerized}
H.~A. Drury, D.~C. Van~Essen, C.~H. Anderson, C.~W. Lee, T.~A. Coogan, and
  J.~W. Lewis.
\newblock Computerized mappings of the cerebral cortex: a multiresolution
  flattening method and a surface-based coordinate system.
\newblock {\em Journal of cognitive neuroscience}, 8(1):1--28, 1996.

\bibitem{elad2003bending}
A.~Elad and R.~Kimmel.
\newblock On bending invariant signatures for surfaces.
\newblock {\em Pattern Analysis and Machine Intelligence, IEEE Transactions
  on}, 25(10):1285--1295, 2003.

\bibitem{fang2012euclidean}
Haw-ren Fang and Dianne~P O'Leary.
\newblock Euclidean distance matrix completion problems.
\newblock {\em Optimization Methods and Software}, 27(4-5):695--717, 2012.

\bibitem{hochbaum1985best}
D.~S. Hochbaum and D.~B. Shmoys.
\newblock A best possible heuristic for the k-center problem.
\newblock {\em Mathematics of operations research}, 10(2):180--184, 1985.

\bibitem{kimmel1998computing}
R.~Kimmel and J.~A. Sethian.
\newblock Computing geodesic paths on manifolds.
\newblock {\em Proceedings of the National Academy of Sciences},
  95(15):8431--8435, 1998.

\bibitem{kohonen1998self}
T.~Kohonen.
\newblock The self-organizing map.
\newblock {\em Neurocomputing}, 21(1):1--6, 1998.

\bibitem{3dor.20151064}
Z.~Lian, J.~Zhang, S.~Choi, H.~ElNaghy, J.~El-Sana, T.~Furuya, A.~Giachetti,
  R.~A. Guler, L.~Lai, C.~Li, H.~Li, F.~A. Limberger, R.~Martin, R.~U.
  Nakanishi, A.~P. Neto, L.~G. Nonato, R.~Ohbuchi, K.~Pevzner, D.~Pickup,
  P.~Rosin, A.~Sharf, L.~Sun, X.~Sun, S.~Tari, G.~Unal, and R.~C. Wilson.
\newblock {Non-rigid 3D Shape Retrieval}.
\newblock In I.~Pratikakis, M.~Spagnuolo, T.~Theoharis, L.~Van Gool, and
  R.~Veltkamp, editors, {\em Eurographics Workshop on 3D Object Retrieval}. The
  Eurographics Association, 2015.

\bibitem{lipman2010biharmonic}
Yaron Lipman, Raif~M Rustamov, and Thomas~A Funkhouser.
\newblock Biharmonic distance.
\newblock {\em ACM Transactions on Graphics (TOG)}, 29(3):27, 2010.

\bibitem{liu2006sub}
R.~Liu, V.~Jain, and H.~Zhang.
\newblock Sub-sampling for efficient spectral mesh processing.
\newblock In {\em Advances in Computer Graphics}, pages 172--184. Springer,
  2006.

\bibitem{mahoney2009cur}
M.~W. Mahoney and P.~Drineas.
\newblock {CUR} matrix decompositions for improved data analysis.
\newblock {\em Proceedings of the National Academy of Sciences},
  106(3):697--702, 2009.

\bibitem{Panozzo:2013aa}
Daniele Panozzo, Ilya Baran, Olga Diamanti, and Olga Sorkine-Hornung.
\newblock Weighted averages on surfaces.
\newblock {\em ACM Transactions on Graphics (TOG)}, 32(4):60 
  2013.

\bibitem{pinkall1993computing}
Ulrich Pinkall and Konrad Polthier.
\newblock Computing discrete minimal surfaces and their conjugates.
\newblock {\em Experimental mathematics}, 2(1):15--36, 1993.

\bibitem{platt2005fastmap}
John~C Platt.
\newblock Fastmap, metricmap, and landmark mds are all {Nystr{\"o}m}
  algorithms.
\newblock In {\em Proc. 10th Int. Workshop on Artificial Intelligence and
  Statistics}, pages 261--268, 2005.

\bibitem{pless2003image}
R.~Pless.
\newblock Image spaces and video trajectories: Using isomap to explore video
  sequences.
\newblock In {\em ICCV}, volume~3, pages 1433--1440, 2003.

\bibitem{roweis2000nonlinear}
S.~T. Roweis and L.~K. Saul.
\newblock Nonlinear dimensionality reduction by locally linear embedding.
\newblock {\em Science}, 290(5500):2323--2326, 2000.

\bibitem{rubner2000perceptual}
Y.~Rubner and C.~Tomasi.
\newblock {\em Perceptual metrics for image database navigation}.
\newblock Springer, 2000.

\bibitem{scholkopf1998nonlinear}
B.~Sch{\"o}lkopf, A.~Smola, and K.~M{\"u}ller.
\newblock Nonlinear component analysis as a kernel eigenvalue problem.
\newblock {\em Neural computation}, 10(5):1299--1319, 1998.

\bibitem{schwartz1989numerical}
E.~L. Schwartz, A.~Shaw, and E.~Wolfson.
\newblock A numerical solution to the generalized mapmaker's problem:
  flattening nonconvex polyhedral surfaces.
\newblock {\em Pattern Analysis and Machine Intelligence, IEEE Transactions
  on}, 11(9):1005--1008, 1989.

\bibitem{schweitzer2001template}
H.~Schweitzer.
\newblock Template matching approach to content based image indexing by low
  dimensional euclidean embedding.
\newblock In {\em Computer Vision, 2001. ICCV 2001. Proceedings. Eighth IEEE
  International Conference on}, volume~2, pages 566--571. IEEE, 2001.

\bibitem{shamai2015classical}
Gil Shamai, Yonathan Aflalo, Michael Zibulevsky, and Ron Kimmel.
\newblock Classical scaling revisited.
\newblock In {\em Proceedings of the IEEE International Conference on Computer
  Vision}, pages 2255--2263, 2015.

\bibitem{shamai2017geodesic}
Gil Shamai and Ron Kimmel.
\newblock Geodesic distance descriptors.
\newblock In {\em Proceedings of the IEEE Conference on Computer Vision and
  Pattern Recognition}, pages 6410--6418, 2017.

\bibitem{shamai2015accelerating}
Gil Shamai, Michael Zibulevsky, and Ron Kimmel.
\newblock Accelerating the computation of canonical forms for 3d nonrigid
  objects using multidimensional scaling.
\newblock In {\em Proceedings of the 2015 Eurographics Workshop on 3D Object
  Retrieval}, pages 71--78. Eurographics Association, 2015.

\bibitem{silva2002global}
V.~D Silva and J.~B. Tenenbaum.
\newblock Global versus local methods in nonlinear dimensionality reduction.
\newblock In {\em Advances in neural information processing systems}, pages
  705--712, 2002.

\bibitem{stein2018natural}
Oded Stein, Eitan Grinspun, Max Wardetzky, and Alec Jacobson.
\newblock Natural boundary conditions for smoothing in geometry processing.
\newblock {\em ACM Transactions on Graphics (TOG)}, 37(2):23, 2018.

\bibitem{surazhsky2005fast}
Vitaly Surazhsky, Tatiana Surazhsky, Danil Kirsanov, Steven~J Gortler, and
  Hugues Hoppe.
\newblock Fast exact and approximate geodesics on meshes.
\newblock In {\em ACM transactions on graphics (TOG)}, volume~24, pages
  553--560. Acm, 2005.

\bibitem{tenenbaum2000global}
Joshua~B Tenenbaum, Vin De~Silva, and John~C Langford.
\newblock A global geometric framework for nonlinear dimensionality reduction.
\newblock {\em Science}, 290(5500):2319--2323, 2000.

\bibitem{xin2012constant}
Shi-Qing Xin, Xiang Ying, and Ying He.
\newblock Constant-time all-pairs geodesic distance query on triangle meshes.
\newblock In {\em Proceedings of the ACM SIGGRAPH symposium on interactive 3D
  graphics and games}, pages 31--38. ACM, 2012.

\bibitem{yu2012isomap}
H.~Yu, X.~Zhao, X.~Zhang, and Y.~Yang.
\newblock {ISOMAP} using {Nystr{\"o}m} method with incremental sampling.
\newblock {\em Advances in Information Sciences \& Service Sciences}, 4(12),
  2012.

\bibitem{yu2009nonlinear}
Kai Yu, Tong Zhang, and Yihong Gong.
\newblock Nonlinear learning using local coordinate coding.
\newblock In {\em Advances in neural information processing systems}, pages
  2223--2231, 2009.

\bibitem{zhou2013locality}
Yin Zhou and Kenneth~E Barner.
\newblock Locality constrained dictionary learning for nonlinear dimensionality
  reduction.
\newblock {\em Signal Processing Letters, IEEE}, 20(4):335--338, 2013.

\bibitem{zigelman2002texture}
Gil Zigelman, Ron Kimmel, and Nahum Kiryati.
\newblock Texture mapping using surface flattening via multidimensional
  scaling.
\newblock {\em Visualization and Computer Graphics, IEEE Transactions on},
  8(2):198--207, 2002.

\end{thebibliography}
\bibliographystyle{plain}
}
\begin{wrapfigure}{l}{0.4\columnwidth}
\includegraphics[width=0.4\columnwidth]{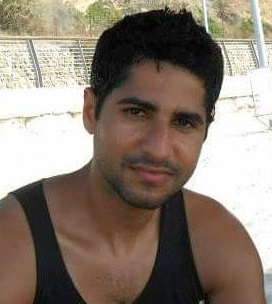}
\end{wrapfigure}
Gil Shamai is a PhD candidate at the Geometric Image Processing (GIP) laboratory, 
 at the Technion---Israel Institute of Technology. 
Shamai is a veteran of the Technion undergraduate Excellence Program and his research was recognized by being the recipient of several awards, including the Meyer Prize, Finzi Prize, and Sherman Interdisciplinary Graduate School Fellowship.
His main interests are theoretical and computational methods in metric geometry and their application to problems in manifold learning, multidimensional scaling procedures and fast geodesics computation, as well as medical imaging and machine learning.

\begin{wrapfigure}{l}{0.4\columnwidth}
\includegraphics[width=0.4\columnwidth]{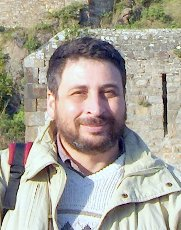}
\end{wrapfigure}
Michael Zibulevsky received his MSc in Electrical Engineering from MIIT---Transportation
Engineering Institute, Moscow and DSc in Operations Research from the Technion---Israel
Institute of Technology (1996). He is currently with the Computer Science Department at
the Technion. Dr. Zibulevsky is one of the original developers of Sparse Component Analysis.
His research interests include numerical methods of nonlinear optimization, sparse signal
representations, deep neural networks and their applications in signal / image processing and
inverse problems.

\begin{wrapfigure}{l}{0.4\columnwidth}
\includegraphics[width=0.4\columnwidth]{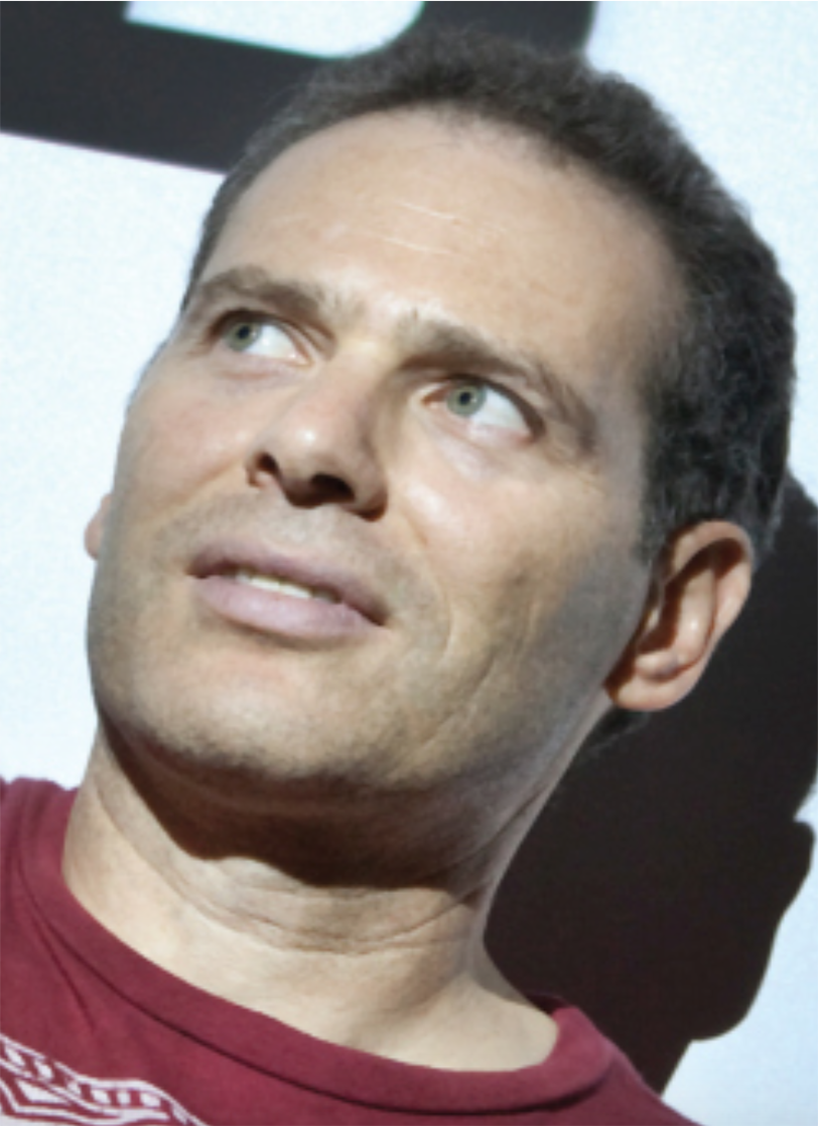}
\end{wrapfigure}
Ron Kimmel is a Professor of Computer Science at the Technion where he holds the Montreal
Chair in Sciences. 
He has worked in various areas of image processing and analysis in computer vision,
image processing, and computer graphics. 
Kimmel's interest in recent years has been shape reconstruction, analysis and learning, 
medical imaging and computational biometry, and applications of metric and differential geometries. 
He is an IEEE Fellow, recipient of the Helmholtz Test of Time Award, and the 
SIAG on Imaging Science Best Paper Prize.
At the Technion he founded and heads the Geometric Image Processing (GIP) Laboratory. He also served as the Vice Dean for Teaching Affairs, and Vice Dean for Industrial Relations. Currently, he serves as Head of Academic Affairs of the Technion's graduate interdisciplinary Autonomous Systems Program (TASP).
Since the acquisition of InVision (a company he co-founded) by Intel in 2012, 
 he has also been heading a small R\&D team as part of Intel's RealSense. 

\end{document}